\documentclass[journal=jpccck,manuscript=article]{achemso}
\usepackage[version=3]{mhchem} 

\usepackage{amsmath,amssymb}
\usepackage{amsfonts}
\usepackage{bm}
\usepackage{stackengine}
\stackMath
\usepackage{natbib}
\usepackage{graphicx}
\usepackage{array}
\usepackage{dcolumn}
\usepackage{amssymb}
\usepackage{multirow}
\usepackage{mathtools}
\newcolumntype{z}[1]{D{.}{.}{#1}}
\usepackage{xcolor}

\usepackage{float}
\usepackage{mciteplus}
\usepackage[T1]{fontenc}
\usepackage[utf8]{inputenc}
\mciteErrorOnUnknownfalse 
\usepackage{tabularx}
\usepackage{rotating}
\usepackage{placeins}
\usepackage{romannum}
\usepackage{physics}

\usepackage{hyperref}
\hypersetup{colorlinks,allcolors=black}

\title{Mechanistic Understanding of Entanglement and Heralding in Cascade Emitters}
\date{\today}
\author{K.~Nasiri~Avanaki}
\affiliation{ Department of Chemistry, Northwestern University, 2145 Sheridan Road, Evanston IL 60208-3113,USA}
\alsoaffiliation{ Department of Chemistry and Chemical Biology, Harvard University, 12 Oxford Street, Cambridge MA 02138, USA}
\author{George C. Schatz}
\affiliation{ Department of Chemistry, Northwestern University, 2145 Sheridan Road, Evanston IL 60208-3113,USA}
\email{g-schatz@northwestern.edu}

\begin{document}
\pagenumbering{arabic}
\maketitle

\begin{abstract}
Semiconductor quantum light sources are
favorable for a wide range of quantum photonic tasks, particularly quantum computing and quantum information processing. Here we theoretically investigate the properties of quantum emitters (QEs) as a source of entangled photons with practical quantum properties including heralding of on-demand single photons. Through the theoretical analysis, we characterize the properties of a cascade (biexciton) emitter, including $(1)$ studies of single-photon purity, 
$(2)$ investigating the first- and second- order correlation functions, and $(3)$ determining the Schmidt number of the entangled photons. The analytical expression derived for the Schmidt number of the cascade emitters reveals a strong dependence on the ratio of decay rates of the first and second photons. Looking into the joint spectral density of the generated biphotons, we show how the purity and degree of entanglement are connected to  the production of heralded single photons.

Our model is further developed to include polarization effects,  fine structure splitting, and the emission delay between the exciton
and biexciton emission. The extended model offers more details about  the underlying mechanism of entangled photon production, and it provides additional degrees of freedom  for manipulating the  system  and characterizing  purity  of the  output  photon. 
The theoretical investigations and the analysis provide a cornerstone for the experimental design and engineering of on-demand single photons.


\end{abstract}
\newpage
Quantum light sources including on-demand single-photon sources are promising candidates in numerous frontier photonic and quantum technologies\cite{Utzat_Science2019,Cosacchi_PRL2019,Krieg_Acs2018,OBrien_NatPh2009}.
Embracing both experimental and theoretical investigations, single-photon emitters (SPE) as intrinsic building blocks play an essential role in quantum
communication\cite{ZhangNatpho2019}, quantum computing\cite{OBrien_NatPh2009} and quantum information processing\cite{Takeda_APL2019}.
In this area, a great deal of work has been done to characterize quantum light  production using
nonlinear crystals\cite{Kwiat_PRL1997} or atomic systems\cite{Thompson_Science2006}, both of which suffer from
 low photon emission rates and limited scalability. Moreover in a commonly used method of twin-state generation, spontaneous parametric down conversion\cite{Mosley_2008,Horn_PRL2012}, the efficiency is low and depends on an inherently in-determinant emission processes. 
 
Semiconductor quantum dots as non-classical  light  emitters  are  of  particular  interest and more favorable for single photon production\cite{Heindel_Natcom2017,Somaschi_NatPho2016,Schweickert_AppPhyLette2017,Ding_PRL2016} due to their high compatibility with the current semiconductor technology.  It has been proven that these excellent quantum emitters\cite{Heindel_Natcom2017,Somaschi_NatPho2016,Schweickert_AppPhyLette2017,Ding_PRL2016,Schulte_Science2015} can
produce single-photon states with high efficiency that are stimulated both by
optical\cite{Somaschi_NatPho2016} and electrical\cite{Schlehahn_APL2016} excitation. 
Moreover, recent advances in fabrication techniques have paved the road for the production of ideal semiconductor single-photon sources\cite{Heindel_Natcom2017,Aharonovich_NatPhoton2016,Sapienza_Natcom2015,Gschrey_APL2013} showing that they possess significant capabilities for producing indistinguishable single-photons or entangled-photon pairs \cite{Senellart_NatNano2017,Aharonovich_NatPhoton2016}. However the implementation depends upon the scalability of quantum emitters (QEs), so
 the on-demand generation of more complex photonic states is still a
 challenging task.

Generally, the characteristic properties of an ideal single-photon source can be classified into three categories:
(i) single-photon purity in which the field does not accommodate more than one photon. This property is determined by the second-order intensity correlation function $g^{(2)}(0)$ through the Hanbury Brown and Twiss (HBT) experiment \cite{BROWN1956, Migdallbook}.
(ii) indistinguishability, that is related to adjusting the quantum interference of two single-photon wavepackets and 
can be measured via  Hong–Ou–Mandel (HOM) interference\cite{HOM1987,Martino_PRApp2014}.
(iii) brightness in which one measures the probability that each light pulse contains a single photon. This measurement gives us more insight into the information contained in the second-order coherence  related to the photon number probabilities \cite{Migdallbook}.

These three properties have been defined differently depending on the scientific community,  although the essential features behind them are the same: a single-photon source should generate light pulses with no more than one photon, the photon should be in a pure quantum state, and it should be generated as efficiently as feasible. All three of these essential features are determined by the second-order correlation function using different experiments.

In general correlation function measurements in the HBT experiment can be regarded as determining the probability of finding two or more photons in the same pulse. This is one of the parameters used to estimate the quality of the single photon source. 
This is significant on the grounds that high single-photon purity assures the security of quantum communications and minimizes errors in quantum computation and simulation\cite{Broome_Science2013, Spagnolo_Natph2014}.

 Bridging the concepts of an ideal SPE introduced  here and the deficiency of scalable single-photon sources motivates the present work, in which we dig into the underlying mechanism for generation of close-to ideal twin-photon states. 
 As part of this, we map out the characteristic parameters relevant to cascade entangled photon emission and show that single-photons derived from pairs generated by cascade emission can be prepared directly in pure states. In addition, we determine the degree of purity via correlation functions and we provide a Schmidt number analysis. Through these studies, we address major obstacles in the advancement of quantum photonics.

 In our theoretical development, the generation of correlated photon pairs in semiconductor emitters is assumed to take place through a biexciton-exciton radiative cascade\cite{Heindel_Natcom2017,Schweickert_AppPhyLette2017,Ding_PRL2016}. In this process, two electron–hole pairs form a biexciton state that radiatively decays with the emission of two photons as mediated by a single exciton
state being an intermediate. The purity of the system is then 
 limited by correlations within the photon pair as determined by the rates of decay of emission of the first and second photons.
\cite{Utzat_Science2019,Krieg_Acs2018,Loredana_NanoLetters2015}. 
 This analysis is based on the fact that the radiative biexciton cascade in a single semiconductor QD provides a source of entangled photons. \cite{Winik_PRB2017, Trotta_Nanolett2014} 
Starting from the biexcitonic ground state of a pre-excited QD, the first electron-hole recombination leads to emission of one photon, and then the second electron of opposite spin recombines with a hole to give a second photon with opposite polarization. This results in anticorrelation in the polarization of the
emitted photons\cite{Solid-state_Nat2017,HighlyIndis_Nat2017,EP_Generation_2007,HighlyEPpairs_PRA2017}. 

\begin{figure}
\includegraphics[clip,width=0.6\columnwidth]{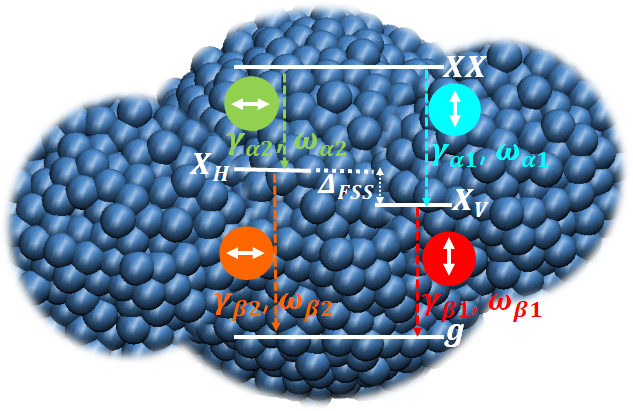}
\caption{Schematic of radiative decay of the biexciton state $\ket{XX}$ in a typical(asymmetric) QD with fine-structure splitting $\Delta_{FSS}$. Here we assume the radiative decay of $\ket{XX}$
generates a pair of vertically or horizontally colinearly polarized photons;  $\dfrac{1}{\sqrt{2}}(\ket{XX_{H}X_{H}}+\ket{XX_{V}X_{V}})$.}
\label{Fig:QDState}
\end{figure}
Our analysis begins with the theoretical prediction just described, but then we progress to a more sophisticated cascade model that is needed when generating entangled photons from semiconductor QDs due to asymmetry in the geometry of the QDs. This imperfectness induces splitting of the intermediate excitonic states, i.e., fine-structure splitting (FSS)  
 ($\Delta_{FSS}$), which is modified as QD size varies.
 This means that we are required to describe the QD biexciton cascade 
using a four-level system composed of the biexciton state ($\ket{XX}$),
two bright intermediate exciton levels ($\ket{X_{H(V)}}$), and a ground state ($\ket{g}$) \cite{Cascade_Semicon_PRL2000}. Spontaneous decay of the biexciton state  
to the ground state thus occurs via two intermediate exciton states leading to
the emission of pairs of photons through the transitions  $\ket{XX}\rightarrow \ket{X_{H(V)}}$ 
and $\ket{X_{H(V)}}\rightarrow\ket{g}$ respectively (see \autoref{Fig:QDState}). 
Thus the intermediate excitonic states lead to spin-dependent properties of the emissions. 
With nonzero-FSS 
\cite{Polarization-corre_PRB2002,EntangledBiexciton_PRB2003}, the degree of entanglement of the entangled polarization photon pairs is lower.  However this can be modified using promising strategies that have been proposed in previous studies\cite{CascadeSemiconNat2006,ManipulatingFSS_APL2007,LowerBoundPRL2010,Entangled_AkopianPRL2006,Cavity-assisted_PRB2009,Pathak_PRB2009}.
\section*{Photon detection and quantum coherence functions in QEs}
 The essence of the HBT experiment is to recognize when the detectors are recording a photocurrent, since the detectors use the photoelectric effect to make local field measurements. 
  For one detector, the photon counting rate is defined by a first-order correlation function \cite{Scullybook},
 $
  G^{(1)}(\textbf{r},t) 
   =\langle
  E^{(-)}(\textbf{r},t)
  E^{(+)}(\textbf{r},t)
  \rangle
      $ . Here the $E^{(+/-)}$'s are positive and negative frequency parts of the fields and the detector is positioned at $\bm{r}$.
       For two photons and two detectors, the joint probability of observing one photoionization at
point $\bm{r}_{2}$ between $t_{2}$ and $t_{2}+dt_{2}$ and another one at point $\bm{r}_{1}$ between
$t_{1}$ and $t_{1}+dt_{1}$ with $t_{1}<t_{2}$ is governed by the second-order quantum mechanical correlation function which is  measured in typical multi-photon counting
experiments.

\begin{figure}
\includegraphics[clip,width=0.6\columnwidth]{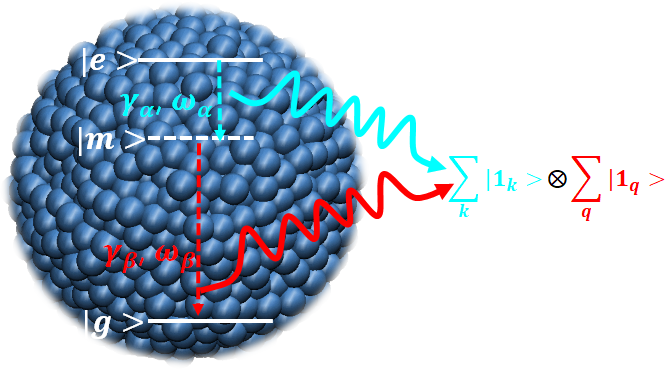}
\caption{ Symmetric (almost perfect) QD source: Three-level configuration used for observation of the two-photon cascade
emission. }
\label{Fig:3Level}
\end{figure}

 In the first part of this work we assume that we have a perfect symmetric QD that can produce highly entangled photons in which the FSS is zero (degenerate intermediate states). All spin and polarization properties are therefore hidden in the notation that we use to describe the states. 
In our cascade emission model, we define the almost perfectly symmetric excited QD as a three-level system where we use $\ket{e}$ and $\ket{m}$ instead of $\ket{XX}$ and $\ket{X_{H(V)}}$ for the excited and the intermediate states respectively. From \autoref{Fig:3Level}, we assume that the system is initially  ($t = 0$) in the top level  $\ket{e}$ with energy $\hbar(\omega_{\alpha}+\omega_{\beta})$ and width $\gamma_{\alpha}$. This means that the lifetimes of the two-photon excited state is $\gamma_{\alpha}^{-1}$.
The first spontaneous emission with frequency $\omega_{k}$ is associated with the transition from $\ket{e}$ to the intermediate state $\ket{m}$ 
 and the second decay is to the ground state $\ket{g}$ via the emission of a photon of
frequency $\omega_{q}$.
  It should be noted that when $\gamma_{\alpha}>\gamma_{\beta}$,
 there will be some population growth in the state $\ket{m}$, but if $\gamma_{\alpha}\ll \gamma_{\beta}$, the state $\ket{m}$  lives for a short period of time and another photon is within a short time delay 
 emitted in the second emission. The latter situation (slow emission followed by fast emission) is the circumstance that we mainly focus on in this work.
\section*{Power Spectrum Analysis: First-order Correlation Function}
 
We assume that at time $t = 0$ the
emitter is in the excited state $\ket{e}$ and the field modes are in the vacuum
state $\ket{0}$.  Given this, the state vector of the particle-field system at time $t$ is described by 
\begin{eqnarray}
  \begin{aligned}
   \ket{\psi(t)}=
   \eta_{e}(t)\ket{e,0}
   +
   \sum_{\textbf{k}}\eta_{m,\textbf{k}}(t)\ket{m,1_{\textbf{k}}}
   +
   \sum_{\textbf{k}\textbf{q}}\eta_{g,\textbf{kq}}(t)\ket{g,1_{\textbf{k}},1_{\textbf{q}}}
   \label{state1}
  \end{aligned}
  \end{eqnarray}
  where the symbol $\ket{1_{\textbf{k}},1_{\textbf{q}}}$ represents the tensor product $\ket{1_{\textbf{k}}}  \otimes \ket{1_{\textbf{q}}}$
of two single photon states of frequency $\omega_{k(q)}$ of subsystem $\alpha(\beta)$ with amplitude of $\eta(\omega_{\textbf{k}},\omega_{\textbf{q}})\equiv \eta_{\textbf{kq}}$.
We determine the states of the particle and 
radiation field as a function of time using the Weisskopf-Wigner approximation where the particle in an excited state decays to the ground state with a
characteristic lifetime but it does not make back and forth transitions.
From the Schr\"odinger equation we have
\begin{eqnarray}
  \begin{aligned}
   \ket{\dot{\psi}(t)}=
   -\dfrac{i}{\hbar}\widehat{H}_{I}
   \ket{\psi(t)}
   \label{SEq}
  \end{aligned}
  \end{eqnarray}
Substituting \autoref{state1} in the Schr\"odinger equation, we arrive at the equations of motion for the amplitudes $\eta_{e}$, $\eta_{m,k}$ and $\eta_{g,k,q}$,
\begin{eqnarray}
  \begin{aligned}
  \dot{\eta}_{e}&=-i\sum_{\textbf{k}}g_{\alpha_{\textbf{k}}}\eta_{m,\textbf{k}}e^{i(\omega_{\alpha}-\omega_{k})t}\\
  \dot{\eta}_{m,\textbf{k}}&=-ig_{\alpha_{\textbf{k}}}\eta_{e}e^{-i(\omega_{\alpha}-\omega_{k})t}
  -i\sum_{\textbf{q}}
  g_{\beta_{\textbf{q}}}\eta_{g,\textbf{kq}}e^{i(\omega_{\beta}-\omega_{q})t}
  \\
  \dot{\eta}_{g,\textbf{kq}}&=
  -i
  g_{\beta_{\textbf{q}}}\eta_{m,\textbf{k}}e^{-i(\omega_{\beta}-\omega_{q})t}
   \label{eqmotion}
  \end{aligned}
  \end{eqnarray}
 We assume that the modes of the field are closely spaced in frequency, so
we replace the summation over $\textbf{k}$ and $\textbf{q}$ by an integral,  $\sum_{\textbf{k}}\rightarrow
  2\dfrac{V}{(2\pi)^{3}}
  \int^{2 \pi}_{0} d\phi
  \int^{ \pi}_{0} \sin \theta d \theta
  \int^{\infty}_{0} k^{2} dk$. Where V is the quantization volume.
The transition dipole; $e\bra{i}\textbf{r}\ket{j}=\boldsymbol\mu_{ij}$ and the radiative decay constants are then defined as
\begin{eqnarray}
  \begin{aligned}
  \gamma_{\alpha}&=\Gamma_{\alpha}/2= \dfrac{1}{4 \pi \epsilon_{0}}
  \dfrac{4 \omega^{3}_{\alpha} \mu^{2}_{em}}{3\hbar c^{3}}\\
  \gamma_{\beta}&=\Gamma_{\beta}/2= \dfrac{1}{4 \pi \epsilon_{0}}
  \dfrac{4 \omega^{3}_{\beta} \mu^{2}_{mg}}{3\hbar c^{3}}
    \end{aligned}
  \end{eqnarray}
  Here $g_{\alpha(\beta)}$ can be taken as a constant associated with the spontaneous emission rate \cite{Scullybook}.
 We then carry out simple integration following \autoref{eqmotion} \cite{AvanakiJPCL2019}, and retrieve the probability amplitudes as
\begin{eqnarray}
  \begin{aligned}
\eta_{m,\textbf{k}}(t)&=-g_{\alpha,\textbf{k}}
\dfrac
{e^{i(\omega_{k}-\omega_{\alpha})t-\gamma_{\alpha}t}-e^{-\gamma_{\beta}t}}
{(\omega_{k}-\omega_{\alpha})+i(\gamma_{\alpha}- \gamma_{\beta})}\\
\eta_{g,\textbf{kq}}(t)&=
  \dfrac{g_{\alpha,\textbf{k}}g_{\beta,\textbf{q}}}{(\omega_{k}-\omega_{\alpha})+i(\gamma_{\alpha}-\gamma_{\beta})}
  \left \{
  \dfrac{1-e^{-\gamma_{\beta}t+i(\omega_{q}-\omega_{\beta})t}}{\omega_{q}-\omega_{\beta}+i\gamma_{\beta}}    - \dfrac{1-e^{i(\omega_{k}+\omega_{q}-\omega_{\alpha}-\omega_{\beta})t-\gamma_{\alpha}t}}{\omega_{k}+\omega_{q}-\omega_{\alpha}-\omega_{\beta}+i\gamma_{\alpha}}
  \right \}
  \label{allcoeff}
\end{aligned}
  \end{eqnarray}
The first emission arises from the transition from $\ket{e}$ to $\ket{m}$, at time $t$,
\begin{eqnarray}
  \begin{aligned}
  -\sum_{k}g_{\alpha} 
  \dfrac{\left \{
  e^{i(\omega_{k}-\omega_{\alpha})t-\gamma_{\alpha}t}
  -e^{-\gamma_{\beta}t}
  \right \}}{(\omega_{k}-\omega_{\alpha})+i(\gamma_{\alpha}-\gamma_{\beta})}
  \ket{m} \otimes \ket{1:\omega_{k},\alpha}
    \end{aligned}
  \end{eqnarray}
 This photon has a Lorentzian distribution in frequency with the width  $|\gamma_{\alpha}-\gamma_{\beta}|$. If  $\gamma_{\alpha}\ll \gamma_{\beta}$, then the state does not
stay for a long time, the second photon is emitted
soon such that the state is given by
 \begin{eqnarray}
  \begin{aligned}
  \sum_{k,q} 
  \dfrac{g_{\alpha}g_{\beta}}{(\omega_{k}-\omega_{\alpha})+i(\gamma_{\alpha}-\gamma_{\beta})}
  \left \{
  \dfrac{1-e^{-\gamma_{\beta}t+i(\omega_{q}-\omega_{\beta})t}}{\omega_{q}-\omega_{\beta}+i\gamma_{\beta}}    - \dfrac{1-e^{i(\omega_{k}+\omega_{q}-\omega_{\alpha}-\omega_{\beta})t-\gamma_{\alpha}t}}{\omega_{k}+\omega_{q}-\omega_{\alpha}-\omega_{\beta}+i\gamma_{\alpha}}
  \right \}\\
  \ket{g}\otimes
  \ket{1_{k},\alpha;1_{q},\beta}
  \label{II_Cas}
  \end{aligned}
  \end{eqnarray}
  Defining  $\mathcal{N}_{EP}$ as the coefficient of the two-photon state ; $\mathcal{N}_{EP}=g_{\alpha}g_{\beta}=\dfrac{2c\sqrt{\gamma_{\alpha} \gamma_{\beta}}}{V^{1/3}}$, we may simplify the two-photon state in the long time limit (times long compared to the radiative decay, $t\gg \gamma_{\alpha}^{-1},\gamma_{\beta}^{-1} $):
\begin{eqnarray}
  \begin{aligned}
  \ket{2P_{cas}}&=\sum_{kq}\eta^{cas}_{\textbf{k},\textbf{q}} \ket{1_{k},\alpha;1_{q},\beta}
  \\
  \eta^{cas}_{\textbf{k},\textbf{q}}&=
  \dfrac{\mathcal{N}_{EP}}{
  {(\omega_{q}-\omega_{\beta}+i\gamma_{\beta}) (\omega_{k}+\omega_{q}-\omega_{\alpha}-\omega_{\beta}+i\gamma_{\alpha}})}
  \label{II_Cas1L}
  \end{aligned}
  \end{eqnarray}
It should be noted that in this limit, both $\eta_{m,\textbf{k}}(t)$ and $\eta_{e}(t)$ are zero and only $\eta^{cas}_{\textbf{k},\textbf{q}}\equiv\eta_{g,\textbf{k},\textbf{q}}(\infty)$ survives which is known as the "joint spectral amplitude".

In order to visualize the  spectra  of the emitted  photons, here we  look into the one-photon correlation function $G^{(1)}(\tau)$, and its normalized counterpart $g^{(1)}(\tau )$,  which gives the degree of first-order temporal coherence between the emission
fields at time $t$ and $t+\tau$  and takes the values $0 \le |g^{(1)}(\tau )| \le 1$ for all light sources \cite{Loudonbook, Scullybook}.
\begin{eqnarray}
  \begin{aligned}
  G^{(1)}_{m}(\tau) 
   &=\langle 
  E^{(-)}_{m}(t)
  E^{(+)}_{m}(t+\tau)
  \rangle\\
  g^{(1)}_{m}(\tau) 
   &=\dfrac{\langle 
  E^{(-)}_{m}(t)
  E^{(+)}_{m}(t+\tau)
  \rangle}
  {\langle E^{(-)}_{m}(t)E^{(+)}_{m}(t)\rangle}
    \quad \quad m=\alpha, \beta
  \end{aligned}
  \end{eqnarray}
In the limit when the line width ($\gamma_{\alpha(\beta)}$) goes to zero, the light field is perfectly coherent and $g
^{(1)}(\infty) = 1$. Using the above expression, the first-order correlation for a linear polarized field $
    E^{(+)}(\textbf{r},t)=\sum_{k} \hat a_\textbf{k}
  e^{-i \omega_{k} t+i\textbf{k}\cdot \textbf{r}}$ is obtained as;
  \begin{eqnarray}
  \begin{aligned}
  G^{(1)}_{\alpha}(\tau)=e^{i\omega_{\alpha}\tau -(\gamma_{\alpha}+\gamma_{\beta})|\tau|},\quad \quad \quad 
  G^{(1)}_{\beta}(\tau)=e^{i\omega_{\beta}\tau -\gamma_{\beta}|\tau|}
  \end{aligned}
  \end{eqnarray}
However, an important property of the first-order correlation function is that it forms a Fourier transform pair with the power spectrum expressed as: $S(\omega)=\dfrac{1}{\pi} Re \int^{\infty} _{0}d \tau G^{(1)}(\tau) e^{-i\omega \tau}$.
The spectrum is obtained by performing a photon number measurement for a specific mode on a given state, i.e. for a given two-photon state density $\rho$, its spectrum is given by $ S(\omega)=\text{Tr}[\hat a^{\dagger}(\omega)\hat a(\omega)\rho]$.
Using \autoref{allcoeff}, we arrive at
\begin{eqnarray}
  \begin{aligned}
  G^{(1)}_{\alpha}(\omega) & = \sum_{k}|\eta_{m}(\omega,\omega_{k})|^{2}
  \\
  G^{(1)}_{\beta}(\omega) &= \sum_{k}|\eta_{g}(\omega_{k},\omega)|^{2}
    \end{aligned}
  \end{eqnarray}
 Hence, the distribution of the emitted photons in the frequency domain is the power spectrum, expressed as
\begin{eqnarray}
  \begin{aligned}
  S_{\alpha}(\omega)&
  =
  \dfrac{2c}{V^{1/3}}
  \dfrac{\gamma_{\alpha}+\gamma_{\beta}}{(\omega-\omega_{\alpha})^{2}+(\gamma_{\alpha}+\gamma_{\beta})^{2}}
  \\
 S_{\beta}(\omega)&
 =
  \dfrac{2c}{V^{1/3}}
 \dfrac{\gamma_{\beta}}{(\omega-\omega_{\beta})^{2}+\gamma_{\beta}^{2}}
  \end{aligned}
  \end{eqnarray}
\begin{figure*}[h]
\begin{center}
\includegraphics[clip,width=0.6\columnwidth]{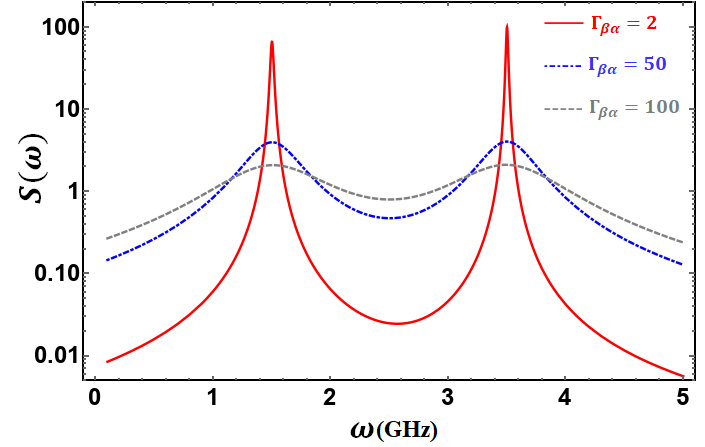}
\caption{
Total spectral density in cascade fluorescence (\autoref{SpecTot}). In these set of calculations we assume that $\gamma_{\alpha}=0.005~\text{GHz}$, $\omega_{\alpha}=1.5~\text{GHz}$ and $\omega_{\beta}=3.5~\text{GHz}$, as described in the text. Different colors show how the spectra and density change as the ratio of $\Gamma_{\beta \alpha}=\dfrac{\gamma_{\beta}}{\gamma_{\alpha}}$ varies. }
\label{Fig:Density_Spec}
\end{center}
\end{figure*}

We see that the 
distribution $S_{\alpha}(\omega)$ associated with the first photon is given by a Lorentzian centered at $\omega_{\alpha}$ and the width at
half-maximum is the sum of the natural widths; $2(\gamma_{\alpha}+\gamma_{\beta})$. In the same fashion, the power spectrum of the second photon is given by $S_{\beta}(\omega)$, leading to a Lorentzian curve of width at half-maximum $2\gamma_{\beta}$, localized around  $\omega_{\beta}$.  These bandwidths are measures of
the coherence time of the emissions.

Note that for a general pure two-photon state $ \ket{2P_{cas}}=\sum_{kq}\eta(\omega_{k},\omega_{q}) \ket{1:\omega_{k},\alpha;1:\omega_{q},\beta}$, we define  the spectral density based on summing up the lineshapes of the photons:
  $S_{2EP}(\omega)=\sum_{k} |\eta(\omega_{k},\omega)|^{2}+|\eta(\omega,\omega_{k})|^{2}$. 
  This means, the emitted photon power spectrum is only determined by the diagonal elements of the corresponding density matrix. This also indicates that the spectrum of a pure two-photon state is the same as that from the corresponding diagonal density matrix. 
 
 Therefore the total spectral density is a blend of two Lorentzian functions with central frequencies $\omega_{\alpha}$, $\omega_{\beta}$
and widths $\gamma_{\alpha}+\gamma_{\beta}$ and $\gamma_{\beta}$ respectively.
\begin{eqnarray}
  \begin{aligned}
  S_{2EP}(\omega)=
  \dfrac{2c}{V^{1/3}}
  \Big[
  \dfrac{\gamma_{\alpha}+\gamma_{\beta}}{(\omega-\omega_{\alpha})^{2}+(\gamma_{\beta}+\gamma_{\alpha})^{2}}
  +
  \dfrac{\gamma_{\beta}}{(\omega-\omega_{\beta})^{2}+\gamma_{\beta}^{2}}
  \Big]
  \label{SpecTot}
  \end{aligned}
  \end{eqnarray}
  In \autoref{Fig:Density_Spec}, we plot the spectral density for cascade fluorescence (right panel).
  Different colors show how the spectra change as the ratio of $\gamma_{\beta}/\gamma_{\alpha}$ varies.
The structure of  the emission spectrum and the tail behavior of the density is in agreement with a previous work \cite{2018acsomega}.  The bimodal shape of the spectrum  originates from overlap of the two emissions as can be explained based on the contribution of each state according to its density of states.
In this figure, and elsewhere in this paper, we performed our calculations using parameters in the same range as polarization-entangled
photon pairs from a biexciton cascade from a single InAs QD embedded in a GaAs/AlAs planar microcavity \cite{CascadeSemiconNJPhys2006,CascadeSemiconNat2006}. In those experiments the pair of entangled photon emissions are at $1.398$~eV and $1.42$~eV. The following references \cite{Polariton-mediatedNat2014,Ebbesen_Angewandte2016,Ebbesen_Angewandte2017,Solid-state_Nat2017} are also relevant.
 \section*{Second-order Correlation Function} 
  Now consider the HBT experiment with a multiphoton source, where we look for the rate of coincidences in the photon-count rates using two detectors. This rate is governed by a second-order correlation function defined as,
  \begin{eqnarray}
  \begin{aligned}
  G^{(2)}(\bm{r},\bm{r'},t,t') =
  \bra{\psi}
  E^{(-)}(\bm{r},t)
  E^{(-)}(\bm{r'},t')
  E^{(+)}(\bm{r'},t')
  E^{(+)}(\bm{r},t)
  \ket{\psi}
  \end{aligned} 
  \end{eqnarray}
  in which the normalized form is 
\begin{eqnarray}
  \begin{aligned}
  g^{(2)}(\bm{r},\bm{r'},t,t') =\dfrac{
  \langle
  E^{(-)}(\bm{r},t)
  E^{(-)}(\bm{r'},t')
  E^{(+)}(\bm{r'},t')
  E^{(+)}(\bm{r},t)
  \rangle
  }
  {
  \langle
  E^{(-)}(\bm{r},t)E^{(+)}(\bm{r},t)
  \rangle
  \langle
  E^{(-)}(\bm{r'},t')E^{(+)}(\bm{r'},t')
  \rangle
  }
  \label{g2origin}
  \end{aligned} 
  \end{eqnarray}
  Here $\ket{\psi}=\sum_{\bm{kq}}\ket{1_{\bm{k}}, 1_{\bm{q}}}$ is the two photon state, and the correlation function refers to detection of photon $\bm{k}$ (at $\bm{r}, t$) followed by detection of photon $\bm{q}$
 (at $\bm{r'}, t'$).
 
In this work, the second-order
correlation function for two-photon emission from cascade emitters can be recast as
\begin{eqnarray}
  \begin{aligned}
  G^{(2)}(t,t') &=
  \bra{2P_{cas}}
  E^{(-)}_{\alpha}(t)
  E^{(-)}_\beta(t')
  E^{(+)}_\beta(t')
  E^{(+)}_{\alpha}(t)
  \ket{2P_{cas}}\\
  &=\sum_{ \{n\} }
  \bra{2P_{cas}}  
  E^{(-)}_{\alpha}(t)E^{(-)}_{\beta}(t')\ket{ \{n\} }
  \bra{\{n\}}
  E^{(+)}_{\beta}(t')E^{(+)}_{\alpha}(t)
  \ket{2P_{cas}}\\
     & =
    \bra{2P_{cas}}
  E^{(-)}_{\alpha}(t)E^{(-)}_{\beta}(t') \ket{0}
  \bra{0}
  E^{(+)}_{\beta}(t')E^{(+)}_{\alpha}(t)
  \ket{2P_{cas}}
  \\
  &=\Psi^{*(2)}(t,t') \Psi^{(2)}(t,t')
  \label{G2}
  \end{aligned} 
  \end{eqnarray}
  Where we defined $\Psi^{(2)}(t,t') \equiv  \bra{0}
  E^{(+)}_{\alpha}(t)E^{(+)}_{\beta}(t')
  \ket{2P_{cas}}$. Note here that a complete set of states; ($\sum_{\{n\}}\ket{\{n\}}
  \bra{\{n\}}=1$) is included.
  Since our two-photon state consists of $\ket{1_{k}, 1_{q}}$  and is annihilated by $E^{(+)}(t)E^{(+)}(t')$,
 only the
 $\ket{0}
  \bra{0}$ term survives, making the final form of \autoref{G2} relatively simple. (It is also true that only the vacuum level persists at long times in the complete wavefunction).
  
  Making use of the two-photon state introduced earlier in \autoref{II_Cas1L} for the detection at times $t$ and $t+\tau$, we then arrive at\cite{Scullybook}
\begin{eqnarray}
  \begin{aligned}
  \Psi^{(2)}(t,t+\tau) 
  &\equiv
  \bra{0}
  E^{(+)}_{\alpha}(t)E^{(+)}_{\beta}(t+\tau)
  \ket{2P_{cas}}\\
  &=\sum_{kq}\eta_{cas} e^{-i\omega_{k}t-i\omega_{q}(t+\tau)}
  \\
  &=
  -\dfrac{V^{1/3}\sqrt{\gamma_{\alpha}\gamma_{\beta}}}{c}
    e^{-(i\omega_{\alpha}+i\omega_{\beta}+\gamma_{\alpha})t}
  \Theta(t)
  e^{-(i\omega_{\beta}+\gamma_{\beta})\tau}
  \Theta(\tau)
  \label{psi2ttau}
   \end{aligned} 
  \end{eqnarray}
Here $\Theta(t)$ is a unit step function.  If we are considering the HBT experimental setup with two detectors ($D_{1}$ and $D_{2}$) for the measurements,  the first term in the above expression indicates that the $\omega_{\alpha}$ photon goes to the $D_{1}$ detector and the second photon $\omega_{\beta}$ to $D_{2}$.  This amplitude should be added to the vice-versa situation to determine the total amplitude,  where the latter has the same form here since both detectors are assumed to be located the same distance from the QD source. 
Substituting the final expression from \autoref{psi2ttau} into \autoref{G2} and ultimately back into \autoref{g2origin}, we see in our bipartite system, the correlations
between parts can be determined with the help of the following (normalized) cross temporal correlation function,
\begin{eqnarray}
  \begin{aligned}
  g^{(2)}_{\times}(t,t+\tau) &=
  \dfrac{\bra{2P_{cas}}
  E^{(-)}_{\alpha}(t)  E^{(-)}_{\beta}(t+\tau)
  E^{(+)}_{\beta}(t+\tau)
  E^{(+)}_{\alpha}(t)
  \ket{2P_{cas}}}
  {\bra{2P_{cas}}  
  E^{(-)}_{\alpha}(t)
  E^{(+)}_{\alpha}(t)
  \ket{2P_{cas}}
  \bra{2P_{cas}}  
  E^{(-)}_{\beta}(t+\tau)
  E^{(+)}_{\beta}(t+\tau)
  \ket{2P_{cas}}
  } 
 \\
  &\approx
  \dfrac{V^{2/3}}{c^{2}}
   \dfrac{{\gamma_{\alpha}\gamma_{\beta}}}{\pi^{2}}
   (\dfrac{\gamma_{\beta}}{\gamma_{\alpha}}-1)
   \Big[
  \Theta(t)     e^{-2\gamma_{\alpha}t}
  \Theta(\tau)  e^{-2\gamma_{\beta}\tau }
  +
  \Theta(-\tau)e^{2(\gamma_{\beta}-\gamma_{\alpha})\tau }
  \Theta(t+\tau)     e^{-2\gamma_{\alpha}t}
  \Big]
    \label{g2xtimetau}
  \end{aligned} 
  \end{eqnarray}
   With regards to the use of entangled photons in quantum photonics, there are always some concerns about pure and reproducible entangled photon generation if the QD is  degraded or if the QD is re-excited after the entangled photons are emitted or
background photons are present. Therefore the purity of the single-photon source is critical for high fidelity QD-photon entanglement, and this generally can be evaluated through the HBT setup, where the following experimentally relevant cross-correlation function is measured,
\begin{eqnarray}
  \begin{aligned}
  g^{(2)}_{\times}(\tau) &=\langle g^{(2)}_{\times}(t,t+\tau)\rangle_{t}=\lim_{T\to\infty}\int^{T}_{-T} g^{(2)}_{\times}(t,t+\tau) dt
  \end{aligned} 
  \end{eqnarray}
  
 Here the total detection time $T$ is taken to be long compared to
the single photon pulsewidth ($T\rightarrow \infty$).
The above formulation 
involves calculating the normalized time dependent second-order correlation function after integrating for a long enough time, $\langle g^{(2)}_{\times}(t,t+\tau)\rangle_{t}$   \cite{Ripka_science2018}. The purity of the system
as a single-photon source is then  extracted from $g^{(2)}_{\times}(\tau=0)$ \cite{Crocker_Express2019,Silva_SciRep2016,Kiraz_PRA2004}.  For our three-level model of Cascade emission the second-order correlation function is found to be
\begin{eqnarray}
  \begin{aligned}
  g^{(2)}_{\times}(\tau) 
  &\approx
  \dfrac{V^{1/3}}{c}
   \dfrac{{\gamma_{\beta}}}{\pi}
   (\dfrac{\gamma_{\beta}}{\gamma_{\alpha}}-1)
      \Big[\Theta(\tau) e^{-2\gamma_{\beta}\tau }+\Theta(-\tau) e^{2\gamma_{\beta}\tau }
   \Big]
   \end{aligned} 
  \end{eqnarray}
Here we see that  $g^{(2)}_{\times}(\tau)$ decays exponentially with $\lvert\tau\rvert$, which makes sense given that the second photon is emitted very shortly after the first. Also, $g^{(2)}_{\times}(\tau=0)$ depends on the radiative decay rates of the two emissions, which is in agreement with the previous studies \citep{Ahn_OptExpress2020,Ripka_science2018,Hamsen_Natphys2018}, and the quantization volume also plays a role   \cite{Ripka_science2018}. When  $\gamma_{\alpha}\ll \gamma_{\beta} $, $g^{(2)}_{\times}(0) >0$, and a positive pure cross correlation is found.
Generally, the area of the peak; $g^{(2)}_{\times}(\tau)$ around
$\tau\sim 0$ (0th peak) gives the normalized coincidence detection probability when two photons are incident in different
inputs of the beam splitter in the HBT experimental setup.  Only in the limit $\gamma_{\alpha}= \gamma_{\beta} $ does this area go to zero. This makes sense as in this limit there is no entanglement (see later discussion of Schmidt numbers).

In this derivation, we have assumed that a light pulse interacts with the system to produce the initial biexciton excited state, and then this decays with no correlation to its initial preparation. However in the actual experimental setup a series of pulses is typically used to excite the system, and there may exist some amplitude from a previous pulse that has not decayed to zero when the next pulse arrives. This leads to a series of peaks in the $g^{(2)}_{\times}(\tau)$ function, but this is not important in the present study. We also note that the inclusion of coherence between the excitation and emission steps will lead to a more complex peak at $\tau=0$, as has often been discussed for other emitters \cite{Crocker_Express2019,Silva_SciRep2016,Kiraz_PRA2004, Yu-Ming_Nat2016}.
  \begin{figure}
    \centering
    \includegraphics[width=.49\textwidth]{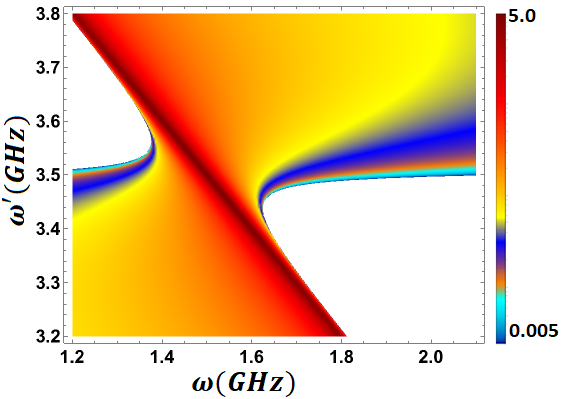}
    \includegraphics[width=.47\textwidth]{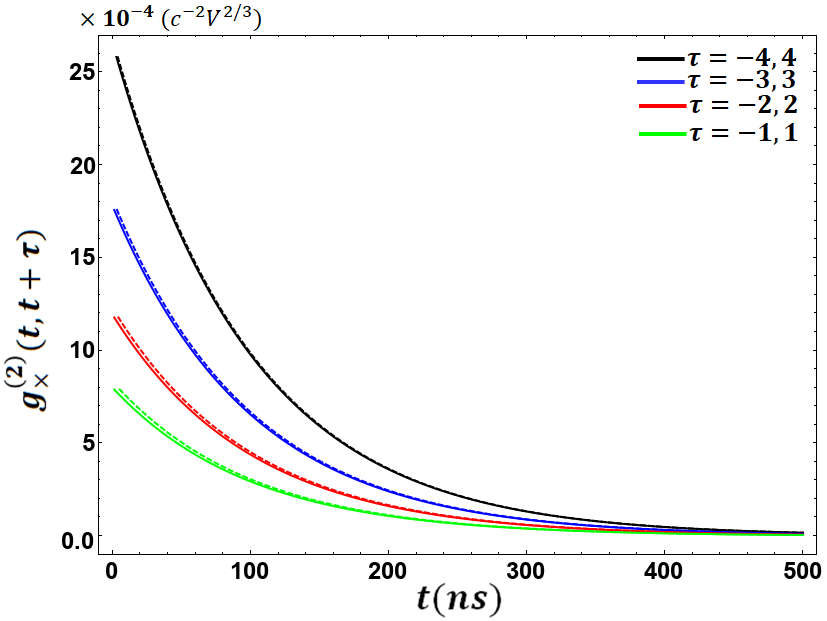}
    \caption{Left: Second-order frequency cross correlation function $g^{(2)}_{\times}(\omega, \omega')$ for the cascade emission process in the frequency domain (logarithmic scale).
    Right: $g^{(2)}_{\times}(t,t+\tau)$, the normalized cross temporal correlation function. Here $\gamma_{\beta}/\gamma_{\alpha}=40$, $\gamma_{\alpha}=0.005~\text{GHz}$, $\omega_{\alpha}=1.5~\text{GHz}$ and $\omega_{\beta}=3.5~\text{GHz}$. }
    \label{fig:g2xwt}
\end{figure}
   As we did in the time domain, in order to fully describe the correlations between two emissions, at different frequencies, $ \omega$ and $ \omega'$, one would have to compute a double Fourier transform according to the cross-correlation  definition; 
   $g^{(2)}_{\times}(\omega, \omega')=\dfrac{1}{\pi^{2}} \Re \int^{\infty} _{-\infty}\int^{\infty} _{-\infty}\,dt \,dt' e^{-i\omega t}e^{-i\omega' t'}g^{(2)}_{\times}(t,t') $.
    This function provides a measure of resemblance of the two photons as a function of the frequency displacement of one relative to the other.  Applying the  cross-correlation definition to \autoref{g2xtimetau}, we obtain:
\begin{eqnarray}
\begin{aligned}
    g^{(2)}_{\times}(\omega,\omega') 
    &\approx
   \dfrac{{\gamma_{\alpha}\gamma_{\beta}}}{\pi^{2}}
   (\dfrac{\gamma_{\beta}}{\gamma_{\alpha}}-1)
   \dfrac{
   4\gamma_{\alpha}\gamma_{\beta}+ (\omega+\omega'-\omega_{\alpha}-\omega_{\beta})(\omega'-\omega_{\beta})
   }
   {[(\omega+\omega'-\omega_{\alpha}-\omega_{\beta})^{2}+4\gamma_{\alpha}^{2}][(\omega'-\omega_{\beta})^{2}+4\gamma_{\beta}^{2}]}
       \label{g2xwt}
\end{aligned}
\end{eqnarray}
  
This function is useful as it determines the width of the frequency anticorrelation associated with the two emitted photons.
 
In Fig.~(\ref{fig:g2xwt}), the $g^{(2)}_{\times}$ function in both time (right) and frequency (left) domains is depicted.
  In the left plot, one observes that the  function
has significant values only on the anti-diagonal, along the line $\omega+\omega'=\omega_{\alpha}+\omega_{\beta}$. This  implies that the corresponding states exhibit strong frequency anticorrelation. The width of the anti-diagonal which gives the characteristic width of
the frequency anticorrelation, 
is equal to $\gamma_{\alpha}$.
  In the right panel of Fig.~(\ref{fig:g2xwt}), $g^{(2)}_{\times}$ from \autoref{g2xtimetau} is plotted versus $t$ for different values of time delay $\tau$ and with $\gamma_{\alpha} <\gamma_{\beta}$. The illustration shows the expected exponential decay of $g^{(2)}_{\times}$ with $t$ as determined by the $\gamma_{\alpha}$ rate, and also that there is exponential decay as a function of $\tau$ as determined by $\gamma_{\beta}$. Note that 
negative anti-correlation is obtained when $\gamma_{\alpha} >\gamma_{\beta}$, which is consistent with the recent study \cite{Ahn_OptExpress2020}.

  \section*{Heralded Single Photons}
So far we have studied the properties of emitted entangled photons and the role of the relevant spectral parameters in the emission spectrum. In this section, we focus on the influence of frequency correlation on the purity of 
heralded single photons that are derived from the two photon state. In the frequency dependence of the general state from a cascade emitter,
the two-photon component 
can be obtained from \autoref{II_Cas1L}  which represents a pure state. Here the joint spectral amplitude generally contains correlations between
frequencies of the sibling photons. As a result of this combination of purity and correlation,
$\ket{2P_{cas}}$ is entangled in the frequency of the two product photons. The purity of either heralded single photon that originates from $\ket{2P_{cas}}$ can then be determined from the 
 density matrix. It is worth recalling that, this property is $inversely$ related to the degree of the entanglement of our two-photon state.
  For a bipartite two-photon source, the density matrix (from \autoref{II_Cas1L}) reads as
 \begin{eqnarray}
  \begin{aligned}
   \rho^{cas}
     &=
   \ket{2P_{cas}}\bra{2P_{cas}}=\sum_{\textbf{kq}}
   |\eta^{cas}_{\textbf{k},\textbf{q}}|^{2} \ket{1_{\textbf{k}},\alpha;1_{\textbf{q}},\beta}
   \bra{1_{\textbf{k}},\alpha;1_{\textbf{q}},\beta}
  \label{rate}
  \end{aligned}
  \end{eqnarray}
where $|\eta^{cas}_{\textbf{k},\textbf{q}}|^{2}$ is the joint spectral probability density.
The purity of either heralded single photon derived from $\ket{2P_{cas}}$ is related to the two reduced density operators of the partner photons, given by:
\begin{eqnarray}
  \begin{aligned}
  \rho_{\alpha}=\text{Tr}_{\beta}\rho= \sum_{k} \xi_{k} \ket{1:\omega_{k}, \alpha}\bra{1:\omega_{k}, \alpha},  
  \quad
  \quad
  \text{where}
  \quad
  \xi_{k}=\sum_{q}\eta_{kq}
  \\
  \rho_{\beta}=\text{Tr}_{\alpha}\rho= \sum_{q} \zeta_{q} \ket{1:\omega_{q}, \beta}\bra{1:\omega_{q}, \beta},  
  \quad
  \quad
  \text{where}
  \quad
  \zeta_{q}=\sum_{k}\eta_{kq}.  
  \end{aligned}
  \end{eqnarray}
 The purity of the individual photons is then determined using
 \begin{eqnarray}
  \begin{aligned}
 \mathcal{P}_{\alpha}=\text{Tr}(\rho^{2}_{\alpha}) ,  
  \quad
  \quad
   \mathcal{P}_{\beta}=\text{Tr}(\rho^{2}_{\beta}). 
   \label{purity1}
  \end{aligned}
  \end{eqnarray}
To see how the spectral correlations in $\ket{2P_{cas}}$ are involved in the purity of the heralded single photons, we examine the Schmidt decomposition of the joint two-photon state. Schmidt decomposition is a characteristic method for describing a bipartite system in terms of a complete set of basis states. Through this decomposition, one can calculate the Schmidt number  which defines the  “degree” of entanglement  of the two-photon state. In this decomposition, \autoref{II_Cas1L} becomes\cite{Eberly_PRL2000,Eberly2006,Chen2017} 
  \begin{eqnarray}
  \begin{aligned}
   \ket{2P_{cas}}= \sum_{k}\sqrt{\lambda_{k}}
  \ket{\phi^{\alpha}_{k}}
  \otimes    
  \ket{\phi^{\beta}_{k}}, 
  \quad\quad
  \text{where}
  \quad
  \bra{\phi^{\mu}_{k}}\ket{\phi^{\mu}_{q}}
  =\delta_{kq},
  \quad
  \text{and}
  \quad
  \sum_{k} \lambda_{k}=1.
  \label{decompose1}
  \end{aligned}
  \end{eqnarray}
The orthonormal basis states $\ket{\phi^{\mu}_{k(q)}}$ ($\mu=\alpha, \beta$ ) are known as Schmidt modes which can be thought of as the basic building
blocks of entanglement in the sense that if the first
photon is described by a function $\ket{\phi^{\alpha}_{k}}$,
we know with certainty that its second sibling is determined
by the corresponding function $\ket{\phi^{\beta}_{k}}$.
Note that each set depends only on one subsystem of $\ket{2P_{cas}}$ and each pair of modes is weighted by its Schmidt magnitude, $\lambda_{k}$. The number of mandatory non-zero components in the sum needed to construct the $\ket{2P_{cas}}$ state (in terms of its Schmidt modes), indicates the effective number of modes that are correlated, while the homogeneity in the distribution of coefficients is determined by the
 Schmidt number, $\kappa$, expressed as 
\begin{eqnarray}
  \begin{aligned}
  \kappa=\dfrac{1}{\text{Tr}(\rho^{2}_{\alpha})}
  =
  \dfrac{1}{\text{Tr}(\rho^{2}_{\beta})}
  =
  \dfrac{1}{\sum_{k=1}^{N}\lambda^{2}_{k}}
  \label{Schnumb1}
  \end{aligned}
  \end{eqnarray}
  Comparing \autoref{Schnumb1} and \autoref{purity1}, we realize that the purity of both reduced states is equal to the sum of the squares of the Schmidt coefficients and thus the inverse of the Schmidt number; $\mathcal{P}_{\alpha(\beta)}= 1/ \kappa$.
 From the experimental point of view, this is relevant to the expected number of  required modes.
 It should also be pointed out that the number of non-zero Schmidt
coefficients in the sum is called the Schmidt rank, or sometimes the dimensionality of the entanglement, as it represents the minimum local Hilbert space dimension required to correctly represent correlations of the quantum state  \cite{Chen2017}.

\section*{Schmidt Analysis}
We now provide an analytical determination of the Schmidt number for the cascade emitter. Different  numerical and analytical frameworks for the Schmidt decomposition of paired photons have been proposed \cite{Eberly_PRL2000,Lamata_2005, Eberly2006,Chen2017} which we can use to advantage for characterizing separability/purity of our bipartite two-photon state. 
To describe the cascade emission presented by $\ket{2P_{cas},\alpha \beta}=\sum_{k,q}\eta(\omega_{k},\omega_{q})\ket{1:\omega_{k},\alpha;1:\omega_{q},\beta}$, we introduce $
  \ket{\psi_{k}}\propto \sum_{m}\eta(\omega_{m},\omega_{k}) \ket{1:\omega_{m}}
  $ as a basis set in which we should find the normalization factor first ( i.e. $\bra{\psi_{j}}\ket{\psi_{k}}=\sum_{m}\eta(\omega_{m},\omega_{k})\eta^{*}(\omega_{m},\omega_{j})$). We then construct the reduced density matrix.
   In our mathematical approach, we approximate the normalized $\ket{\psi_{k}}$ as a piece-wise state $\ket{\phi_{s}}$ by discretizing $\omega_{k}$ into a small interval of $2 \gamma_{\alpha}$, i.e. $\omega_{k}= 2 N_{s}\gamma_{\alpha}$ where $N_{s}$ is an integer. 
  The normalized basis set is therefore defined as
  \begin{eqnarray}
  \begin{aligned}
  \ket{\phi_{s}}&\approx 
  \sqrt{\dfrac{\pi}{\gamma_{\beta}}
\dfrac  {(2\gamma_{\alpha} N_{s}-\omega_{\beta})^{2}+\gamma_{\beta}^{2} }
  {4\gamma_{\alpha}^{2}}}
    \int^{2\gamma_{\alpha} (N_{s}+1)}_{2\gamma_{\alpha} N_{s}}\sum_{m}\eta(\omega_{m},\omega)d\omega \ket{1:\omega_{m}} \quad \quad N_{s}=\omega_{k}/2\gamma_{\alpha}
  \end{aligned}
  \end{eqnarray}
 Here we have approximated the sum over 'm' as $\sum_{m} \leftrightarrow \int \dfrac{L}{2\pi c}d \omega_{m}$ and used  Cauchy's integral theorem. Then the orthonormal relations read as $ \bra{\phi_{r}}\ket{\phi_{s}}
  \approx \delta_{rs}$ and  we obtain the reduced density matrix within a very good approximation as (see more details in SI).
\begin{eqnarray}
  \begin{aligned}
   \rho^{red}
    &=\sum_{N_{s}}\dfrac{\gamma_{\beta}}{\pi}
  \dfrac{2\gamma_{\alpha}}
  {(2\gamma_{\alpha} N_{s}-\omega_{\beta})^{2}+\gamma_{\beta}^{2} }
  \ket{\phi_{s}}\bra{\phi_{s}}
  \end{aligned}
  \end{eqnarray}
Accordingly the “Schmidt coefficients” $\{ \sqrt{\lambda_{s}}\}$ are given by the square roots of the eigenvalues of the reduced density matrix, $\sqrt{\lambda_{s}}
   \approx
   \sqrt{\dfrac{\gamma_{\beta}}{\pi}
  \dfrac{2\gamma_{\alpha}}
  {(2\gamma_{\alpha} N_{s}-\omega_{\beta})^{2}+\gamma_{\beta}^{2} }
  }$ 
 and we can simply show that $\sum_{s} \lambda_{s}\approx 1$.
 
 Generally finding an analytical expression for Schmidt modes and Schmidt number is tricky and complicated. However using the purity definition which is the inverse of the Schmidt number, it is possible to obtain the Schmidt number with no further approximation.
 Thus the Schmidt number of the bipartite two photon
state generated by cascade emission can be obtained from \autoref{Schnumb1}. 
The resulting formula after some lengthy algebra is:
\begin{eqnarray}
  \begin{aligned}
  \kappa=\dfrac{1}{\sum_{mn}|\sum_{k}\eta_{km}\eta^{*}_{kn}|^{2}}=1+\dfrac{\gamma_{\beta}}{\gamma_{\alpha}}
   \end{aligned}
  \end{eqnarray}
 We see that the minimum Schmidt number from this analysis falls at unity 
 corresponding to the limit $\gamma_{\alpha}\gg \gamma_{\beta}$.  For this situation, the population
 of the doubly excited state in the three-level QD source  will decay to the intermediate state within a very short time, and then the transition from intermediate level to ground state occurs on a longer time scale. Therefore, the total energy of the two-photon state varies significantly as a function of time, and there is weak frequency anti-correlation. Also, if $\gamma_{\alpha}$ is extremely large,  de-excitation from the top state occurs instantaneously after coupling to the radiation field is turned on, and the second photon is uncorrelated from the first. 
 
 If $\gamma_{\alpha}$ is very small compared with $\gamma_{\beta}$, then the second photon is emitted soon after the first photon emission, and the fluctuations in the total energy of the two-photon state will be negligible. This implies that the two photons have strong frequency anti-correlation, which is consistent with our earlier analysis. In the extreme case where $\gamma_{\alpha}$ is close to zero, the photon pairs are fully frequency-anticorrelated, the expression for $\eta_{kq}$ breaks down  into $\eta(\omega_{k},\omega-\omega_{k})$ , and the state is clearly in the form of a Schmidt decomposition. 
  \begin{figure}
    \centering
        \includegraphics[width=1.05
        \textwidth]{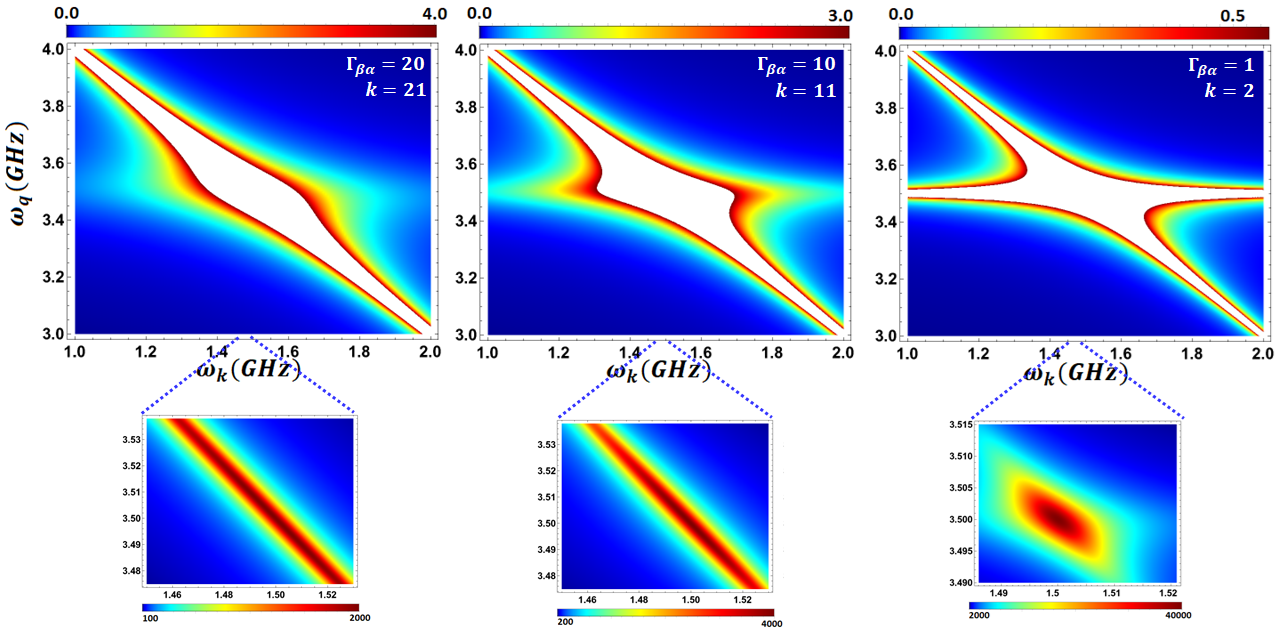}\\
           \caption{Joint spectral density (JSD) of QD emitter in three-level model. JSD profile goes from 
           a broad linewidth ($\gamma_{\alpha}$) to being  
           symmetric along the digonal while the ratio  of $\Gamma_{\beta \alpha}=\gamma_{\beta}/\gamma_{\alpha}$ is increased. Here we assume that $\gamma_{\alpha}=0.005~\text{GHz}$, $\omega_{\alpha}=1.5~\text{GHz}$ and $\omega_{\beta}=3.5~\text{GHz}$.
           }
    \label{fig:JointSpect}
\end{figure}
 Due to the flatness of the distribution of Schmidt coefficients, the corresponding Schmidt number can be very large. From this we conclude that a larger ratio of $\gamma_{\beta}/\gamma_{\alpha}$ gives rise to better frequency correlations i.e. stronger entanglement of the total state. 
 
 In contrast to the high entanglement case, to collect heralded single photons in a pure state, one must ensure that the relevant parameters of the system closely meet the condition; $\gamma_{\alpha}\gg \gamma_{\beta}$. This also means that to separate the entangled photons into two single photons, one should reduce
correlations such that the Schmidt number has a low value. For the cascade source we consider, the only way this can be done is to make the joint spectral density given by $|\eta^{cas}_{\textbf{k},\textbf{q}}|^{2}$ be factorable.

We plot the Joint Spectral Density (JSD) in \autoref{fig:JointSpect}.  This  characterizes the joint spectrum of the two photons, and it can be manipulated
via the emission bandwidths and by other relevant parameters of the two photon state.
   Experimentaly this can be done by measuring the JSD profile with tunable narrow band filters \cite{Valencia_PRL2007} in a same manner as HOM quantum interference  quantifies the two-photon coherence bandwidth and
the indistinguishability of the photon pair. This is a common method for heralding the signal and idler photons in a parametric down conversion experiment.

   The JSDs in \autoref{fig:JointSpect} illustrate how the frequency correlation is related to the degree of entanglement of states of the two-photons when the ratio of $\gamma_{\beta}/\gamma_{\alpha}$ is changed. Here the vertical and horizontal axes show the frequencies of the first and second emissions. The direct consequences of the entanglement are seen when the Schmidt number is decreased (going from the left to the right) over the range $\kappa=21-2$. 
   The results indicate that the probability of frequency correlation is highest in a very short range close to the transition frequencies, $\omega_{\alpha}$ and  $\omega_{\beta}$. Noticeably we see that the distribution is highly aligned with the anti-diagonal wherein $\omega_{\alpha}+\omega_{\beta}= \omega_{k}+\omega_{q}$ when $\kappa$ is high. As $\kappa$ is decreased there is a broadening of the distribution centered on $\omega_{q}=\omega_{\beta}$.
   Also, the correlation intensity is reduced as the JSD profile goes from being closely aligned along the anti-diagonal  to a broadened line shape.  We may conclude that the more asymmetric and spread-out is the spectral density, the less entangled the photons are.

   The images in figure \autoref{fig:JointSpect} also zoom in for a smaller frequency range around the anti-diagonal, and paying closer attention to the middle part of these distributions in which the intensity is very high.
   Overall, we recognize that a biphoton state can be ideally suited for generating heralded pure-state single photons when  the side lobes that hinder the generation of the entangled photon,  symmetrically are enhanced.
   Although the JSD is very informative about the properties of the states, the JSD alone is insufficient  to conclude that they are frequency entangled \cite{Chen2017}.

 \section*{Polarization Effects in Entangled Photons}
\begin{figure*}[h]
\begin{center}
\includegraphics[clip,width=0.95\columnwidth]{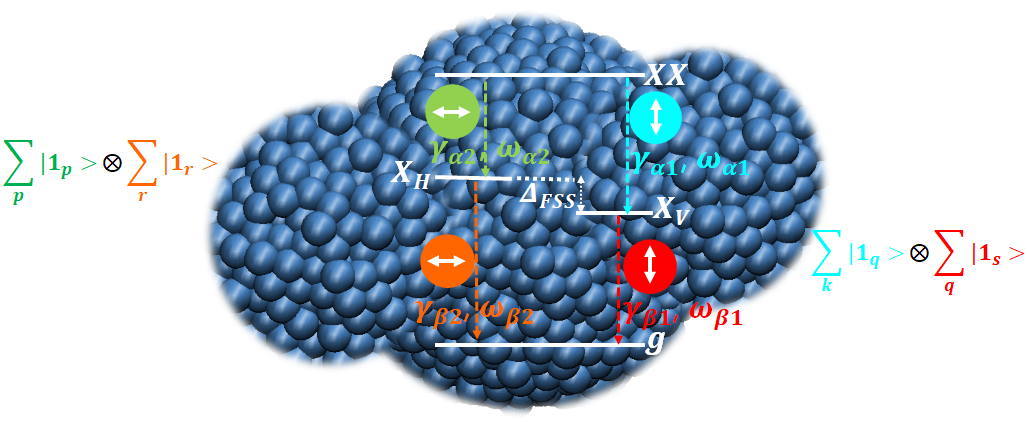}
\caption{ Entangled photon generation from biexciton cascade emission: The final two-photon state is created by sequential emission of two photons (XX and X) separated by a short time delay $\tau$. The resulting state is a superposition of horizontal (H) and vertical (V) polarization states. }
\label{Fig:QD_FSS2}
\end{center}
\end{figure*}
  As we explained in the introduction section,
it is more appropriate if we define a QD biexciton cascade 
using a four-level system composed of a biexciton state (XX),
two bright intermediate exciton levels ($X_{H(V)}$) with different polarizations(either horizontal (H) or vertical (V)) and a ground state (g).
Therefore, the decay proceeds via one of two paths (See \autoref{Fig:QD_FSS2}).
Here we assume that the system is initially 
in a superposition of the biexciton-exciton photonic states. After emitting the first photon, it evolves to the exciton (X) state in which the degeneracy is split. The quantum dot remains in a superposition of $X_{H}$ and $X_{V}$ for a time delay $\tau_{e}$, during which a phase difference develops due to the fine structure splitting $\Delta_{\text{FSS}}$ between different exciton states. Finally, the exciton photon $X_{H(V)}$ with the same polarization as the first biexciton photon is emitted, and the QD goes back to the ground state.
  The system is now found in a superposition of orthogonally polarized photon pair states, with a phase between them that is characterized by the time delay $\tau_{e}$
    (generally in the order of tens of ps). 
   If we denote the state of the photon in each emission by the corresponding state of the QD,
 including for the polarization effect, our cascade two-photon emission wavefunction in \autoref{II_Cas1L} is modified to:
  \begin{eqnarray}
  \begin{aligned}
  \ket{2P_{cas}}&= \dfrac{1}{\sqrt{2}}\Big( 
  \sum_{p,r}\eta^{(H)}_{p,r}\ket{XX_{H}X_{H}} + 
  \sum_{q,s}\eta^{(V)}_{q,s}\ket{XX_{V}X_{V}}
  \Big)   \quad \quad\text{for}\quad   \Delta_{\text{FSS}}=0 
  \\
    \ket{2P_{cas}}&= \dfrac{1}{\sqrt{2}}\Big( 
 \sum_{p,r}\eta^{(H)}_{p,r} \ket{XX_{H}X_{H}} +
 \sum_{q,s}\eta^{(V)}_{q,s}e^{i\Delta_{\text{FSS}} \tau_{e}/\hbar} \ket{XX_{V}X_{V}}
 \Big)
    \quad \quad \text{for}\quad  \Delta_{\text{FSS}}\neq 0
  \label{II_Cas1}
  \end{aligned}
  \end{eqnarray}
  Here
   \begin{eqnarray}
  \begin{aligned}
  \eta^{(H)}_{p,r}&=
  \dfrac{\mathcal{N}_{1}}{
  {(\omega_{r}-\omega_{\beta_{1}}+i\gamma_{\beta_{1}}) (\omega_{p}+\omega_{r}-\omega_{\alpha_{1}}-\omega_{\beta_{1}}+i\gamma_{\alpha_{1}}})}\\
  \eta^{(V)}_{q,s}&=
  \dfrac{\mathcal{N}_{2}}{
  {(\omega_{s}-\omega_{\beta_{2}}+i\gamma_{\beta_{2}}) (\omega_{q}+\omega_{s}-\omega_{\alpha_{2}}-\omega_{\beta_{2}}+i\gamma_{\alpha_{2}}})}
  \end{aligned}
  \end{eqnarray}
   So the joint spectral density is defined as
   \begin{eqnarray}
  \begin{aligned}
 JS_{H-V}&=|\eta^{(H)}_{p,r}+ e^{i\Delta_{\text{FSS}} \tau_{e}/\hbar}\eta^{(V)}_{q,s}|^{2}
   \label{JSI_Pol}
  \end{aligned}
  \end{eqnarray}
   \begin{figure}
    \centering
        \includegraphics[width=1.05\textwidth]{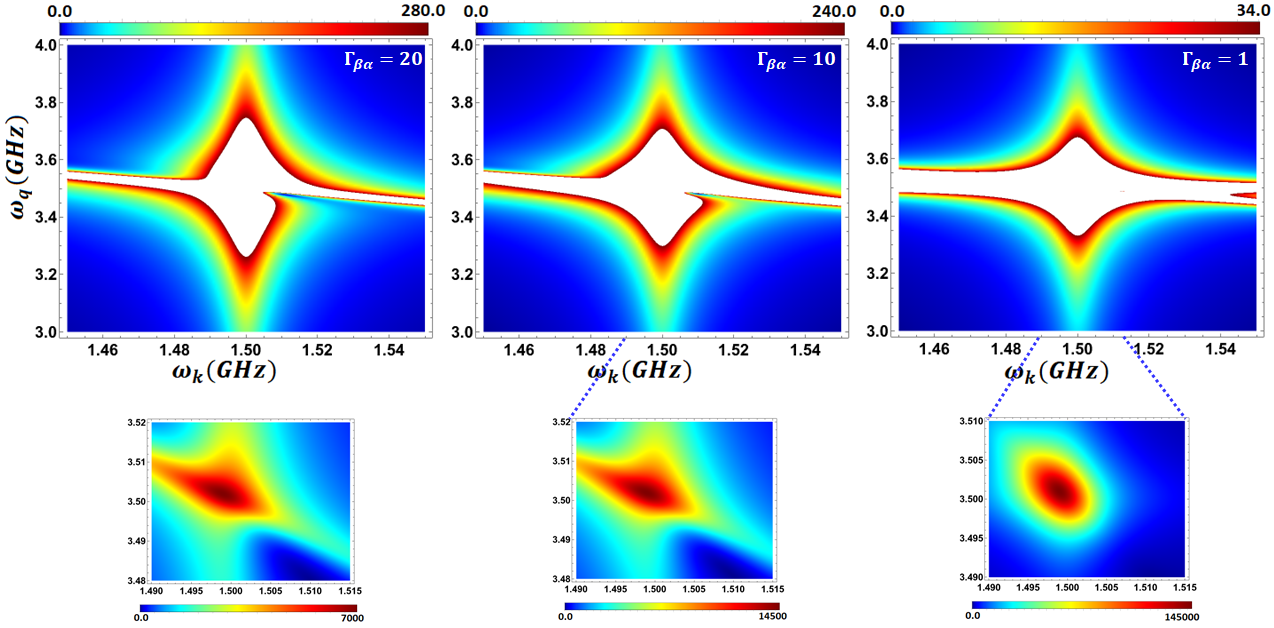}
    \caption{Joint spectral density in cascade emission from typical QD. The polarization of states is included. Different values of $\Gamma_{\beta \alpha}=\dfrac{\gamma_{\beta}}{\gamma_{\alpha}}$ and the phase $\phi=\Delta_{\text{FSS}} \tau_{e}/\hbar=\pi/4$ are used.}
    \label{fig:JointSpect_FSS2}
\end{figure}
  With this perspective, we have more degrees of freedom for choosing the relevant parameters of the system to control purity of the output photons.
  Studies have shown that the maximum entanglement is obtained when the $\omega_{\alpha_{i}}=\omega_{\beta_{j}}$ and $\gamma_{\alpha_{i}}=\gamma_{\beta_{j}}$ ( where $i,j=1,2$) \cite{Hudson_PRL2007, Stevenson_PRL2008,Trotta_Nanolett2014,Winik_PRB2017}. We are mostly interested in predicting the degree of the entanglement by looking at the joint spectral density plotted in \autoref{fig:JointSpect_FSS2}.
  Note that the horizontal axis is relative to the frequency of $X_{H}(_{V})$
in the excitonic transitions and the vertical axis is relevant to the frequency of
$XX_{H}(_{V})$ in the biexcitonic transitions. 
Compared to the three-level model discussed earlier, here the model is closer to reality and we see more details in the JSD plots. More importantly we see the star shaped emission pattern of exciton and biexciton  more explicitly at different frequencies when $\Gamma_{\beta \alpha}$ is higher.
As this ratio becomes smaller from left to  right, the probability density becomes more circular in a narrow domain of frequency around  $\omega_{\alpha}$ and  $\omega_{\beta}$ which assures better separation (greater purity) in the photon production.
Note that the broken symmetry on the right hand plot can be improved by optimizing geometry of the QD experimentally \cite{CascadeSemiconNat2006,ManipulatingFSS_APL2007,LowerBoundPRL2010,Entangled_AkopianPRL2006,Cavity-assisted_PRB2009,Pathak_PRB2009} and removing the phase term, $\phi=\Delta_{\text{FSS}} \tau/\hbar$.
Full analysis and more details of other properties of cascade emitters with this model will be reported in our future work.
  
\section*{Conclusion}
  In conclusion, we theoretically studied the underlying mechanism of entangled two-photon generation in semiconductor QD emitters including use of these emitters as an on-demand single photon source. We developed analytical expressions for the characteristic parameters associated with the first- and second- order correlation function, and the Schmidt number of the entangled cascade emission.
  We extended our model by including for the effects of polarization and  fine structure splitting, and the emission delay of the exciton
relative to the biexciton. The extended model broadens our vision to see the  capacity of other relevant parameters for the practical application of semiconductor quantum dot emitters as single source emitters and offers more details about  the underlying mechanism and purity properties  of entangled photon production. Although we have only investigated this effect in the joint spectral density, it is straightforward to include other properties as well.
  The theoretical studies and the analysis here provides guidelines for the experimental design and engineering of on-demand single photon source applications as diverse as quantum computing and quantum information.
  \begin{suppinfo}
The analytical derivations of Schmidt number for cascade emission is explained here.
\end{suppinfo}

  \begin{acknowledgement}
This work was supported by the U.S. National Science Foundation under Grant No. CHE-1760537. This research was supported in part through the computational resources and staff contributions provided for the Quest high performance computing facility at Northwestern University which is jointly supported by the Office of the Provost, the Office for Research, and Northwestern University Information Technology. 
\end{acknowledgement}
\bibliography{Refs}

\providecommand{\latin}[1]{#1}
\makeatletter
\providecommand{\doi}
  {\begingroup\let\do\@makeother\dospecials
  \catcode`\{=1 \catcode`\}=2 \doi@aux}
\providecommand{\doi@aux}[1]{\endgroup\texttt{#1}}
\makeatother
\providecommand*\mcitethebibliography{\thebibliography}
\csname @ifundefined\endcsname{endmcitethebibliography}
  {\let\endmcitethebibliography\endthebibliography}{}
\begin{mcitethebibliography}{65}
\providecommand*\natexlab[1]{#1}
\providecommand*\mciteSetBstSublistMode[1]{}
\providecommand*\mciteSetBstMaxWidthForm[2]{}
\providecommand*\mciteBstWouldAddEndPuncttrue
  {\def\EndOfBibitem{\unskip.}}
\providecommand*\mciteBstWouldAddEndPunctfalse
  {\let\EndOfBibitem\relax}
\providecommand*\mciteSetBstMidEndSepPunct[3]{}
\providecommand*\mciteSetBstSublistLabelBeginEnd[3]{}
\providecommand*\EndOfBibitem{}
\mciteSetBstSublistMode{f}
\mciteSetBstMaxWidthForm{subitem}{(\alph{mcitesubitemcount})}
\mciteSetBstSublistLabelBeginEnd
  {\mcitemaxwidthsubitemform\space}
  {\relax}
  {\relax}

\bibitem[Utzat \latin{et~al.}(2019)Utzat, Sun, Kaplan, Krieg, Ginterseder,
  Spokoyny, Klein, Shulenberger, Perkinson, Kovalenko, and
  Bawendi]{Utzat_Science2019}
Utzat,~H.; Sun,~W.; Kaplan,~A. E.~K.; Krieg,~F.; Ginterseder,~M.; Spokoyny,~B.;
  Klein,~N.~D.; Shulenberger,~K.~E.; Perkinson,~C.~F.; Kovalenko,~M.~V.
  \latin{et~al.}  Coherent single-photon emission from colloidal lead halide
  perovskite quantum dots. \emph{Science} \textbf{2019}, \emph{363},
  1068--1072\relax
\mciteBstWouldAddEndPuncttrue
\mciteSetBstMidEndSepPunct{\mcitedefaultmidpunct}
{\mcitedefaultendpunct}{\mcitedefaultseppunct}\relax
\EndOfBibitem
\bibitem[Cosacchi \latin{et~al.}(2019)Cosacchi, Ungar, Cygorek, Vagov, and
  Axt]{Cosacchi_PRL2019}
Cosacchi,~M.; Ungar,~F.; Cygorek,~M.; Vagov,~A.; Axt,~V.~M. Emission-Frequency
  Separated High Quality Single-Photon Sources Enabled by Phonons. \emph{Phys.
  Rev. Lett.} \textbf{2019}, \emph{123}, 017403\relax
\mciteBstWouldAddEndPuncttrue
\mciteSetBstMidEndSepPunct{\mcitedefaultmidpunct}
{\mcitedefaultendpunct}{\mcitedefaultseppunct}\relax
\EndOfBibitem
\bibitem[Krieg \latin{et~al.}(2018)Krieg, Ochsenbein, Yakunin, ten Brinck,
  Aellen, Süess, Clerc, Guggisberg, Nazarenko, Shynkarenko, Kumar, Shih,
  Infante, and Kovalenko]{Krieg_Acs2018}
Krieg,~F.; Ochsenbein,~S.~T.; Yakunin,~S.; ten Brinck,~S.; Aellen,~P.;
  Süess,~A.; Clerc,~B.; Guggisberg,~D.; Nazarenko,~O.; Shynkarenko,~Y.
  \latin{et~al.}  Colloidal $CsPbX_{3}$ (X = Cl, Br, I) Nanocrystals 2.0:
  Zwitterionic Capping Ligands for Improved Durability and Stability. \emph{ACS
  Energy Letters} \textbf{2018}, \emph{3}, 641--646\relax
\mciteBstWouldAddEndPuncttrue
\mciteSetBstMidEndSepPunct{\mcitedefaultmidpunct}
{\mcitedefaultendpunct}{\mcitedefaultseppunct}\relax
\EndOfBibitem
\bibitem[O'Brien \latin{et~al.}(2009)O'Brien, Furusawa, and
  Vučković]{OBrien_NatPh2009}
O'Brien,~J.~L.; Furusawa,~A.; Vučković,~J. Photonic quantum technologies.
  \emph{Nature Photonics} \textbf{2009}, \emph{3}, 687\relax
\mciteBstWouldAddEndPuncttrue
\mciteSetBstMidEndSepPunct{\mcitedefaultmidpunct}
{\mcitedefaultendpunct}{\mcitedefaultseppunct}\relax
\EndOfBibitem
\bibitem[Zhang \latin{et~al.}(2019)Zhang, Haw, Cai, Xu, Assad, Fitzsimons,
  Zhou, Zhang, Yu, Wu, Ser, Kwek, and Liu]{ZhangNatpho2019}
Zhang,~G.; Haw,~J.~Y.; Cai,~H.; Xu,~F.; Assad,~S.~M.; Fitzsimons,~J.~F.;
  Zhou,~X.; Zhang,~Y.; Yu,~S.; Wu,~J. \latin{et~al.}  An integrated silicon
  photonic chip platform for continuous-variable quantum key distribution.
  \emph{Nature Photonics} \textbf{2019}, \emph{12}, 839\relax
\mciteBstWouldAddEndPuncttrue
\mciteSetBstMidEndSepPunct{\mcitedefaultmidpunct}
{\mcitedefaultendpunct}{\mcitedefaultseppunct}\relax
\EndOfBibitem
\bibitem[Takeda and Furusawa(2019)Takeda, and Furusawa]{Takeda_APL2019}
Takeda,~S.; Furusawa,~A. Toward large-scale fault-tolerant universal photonic
  quantum computing. \emph{APL Photonics} \textbf{2019}, \emph{4}, 060902\relax
\mciteBstWouldAddEndPuncttrue
\mciteSetBstMidEndSepPunct{\mcitedefaultmidpunct}
{\mcitedefaultendpunct}{\mcitedefaultseppunct}\relax
\EndOfBibitem
\bibitem[Kwiat \latin{et~al.}(1995)Kwiat, Mattle, Weinfurter, Zeilinger,
  Sergienko, and Shih]{Kwiat_PRL1997}
Kwiat,~P.~G.; Mattle,~K.; Weinfurter,~H.; Zeilinger,~A.; Sergienko,~A.~V.;
  Shih,~Y. New High-Intensity Source of Polarization-Entangled Photon Pairs.
  \emph{Phys. Rev. Lett.} \textbf{1995}, \emph{75}, 4337--4341\relax
\mciteBstWouldAddEndPuncttrue
\mciteSetBstMidEndSepPunct{\mcitedefaultmidpunct}
{\mcitedefaultendpunct}{\mcitedefaultseppunct}\relax
\EndOfBibitem
\bibitem[Thompson \latin{et~al.}(2006)Thompson, Simon, Loh, and
  Vuletic]{Thompson_Science2006}
Thompson,~J.; Simon,~J.; Loh,~H.; Vuletic,~V. A high-brightness source of
  narrowband, identical-photon pairs. \emph{Science} \textbf{2006}, \emph{313},
  74--77\relax
\mciteBstWouldAddEndPuncttrue
\mciteSetBstMidEndSepPunct{\mcitedefaultmidpunct}
{\mcitedefaultendpunct}{\mcitedefaultseppunct}\relax
\EndOfBibitem
\bibitem[Mosley \latin{et~al.}(2008)Mosley, Lundeen, Smith, and
  Walmsley]{Mosley_2008}
Mosley,~P.~J.; Lundeen,~J.~S.; Smith,~B.~J.; Walmsley,~I.~A. Conditional
  preparation of single photons using parametric downconversion: a recipe for
  purity. \emph{New Journal of Physics} \textbf{2008}, \emph{10}, 093011\relax
\mciteBstWouldAddEndPuncttrue
\mciteSetBstMidEndSepPunct{\mcitedefaultmidpunct}
{\mcitedefaultendpunct}{\mcitedefaultseppunct}\relax
\EndOfBibitem
\bibitem[Horn \latin{et~al.}(2012)Horn, Abolghasem, Bijlani, Kang, Helmy, and
  Weihs]{Horn_PRL2012}
Horn,~R.; Abolghasem,~P.; Bijlani,~B.~J.; Kang,~D.; Helmy,~A.~S.; Weihs,~G.
  Monolithic Source of Photon Pairs. \emph{Phys. Rev. Lett.} \textbf{2012},
  \emph{108}, 153605\relax
\mciteBstWouldAddEndPuncttrue
\mciteSetBstMidEndSepPunct{\mcitedefaultmidpunct}
{\mcitedefaultendpunct}{\mcitedefaultseppunct}\relax
\EndOfBibitem
\bibitem[Heindel \latin{et~al.}(2017)Heindel, Thoma, von Helversen, Schmidt,
  Schlehahn, Gschrey, Schnauber, Schulze, Strittmatter, Beyer, Rodt, Carmele,
  Knorr, and Reitzenstein]{Heindel_Natcom2017}
Heindel,~T.; Thoma,~A.; von Helversen,~M.; Schmidt,~M.; Schlehahn,~A.;
  Gschrey,~M.; Schnauber,~P.; Schulze,~J.~H.; Strittmatter,~A.; Beyer,~J.
  \latin{et~al.}  A bright triggered twin-photon source in the solid state.
  \emph{Nature Communications} \textbf{2017}, \emph{8}, 14870\relax
\mciteBstWouldAddEndPuncttrue
\mciteSetBstMidEndSepPunct{\mcitedefaultmidpunct}
{\mcitedefaultendpunct}{\mcitedefaultseppunct}\relax
\EndOfBibitem
\bibitem[Somaschi \latin{et~al.}(2016)Somaschi, Giesz, De~Santis, Loredo,
  Hornecker, Portalupi, Grange, Antón, Demory, Gómez, Sagnes,
  Lanzillotti-Kimura, Lemaítre, Auffeves, White, Lanco, and
  Senellart]{Somaschi_NatPho2016}
Somaschi,~N.; Giesz,~V.; De~Santis,~L.; Loredo,~M.~P.,~J. C.and~Almeida;
  Hornecker,~G.; Portalupi,~S.~L.; Grange,~T.; Antón,~C.; Demory,~J.;
  Gómez,~C. \latin{et~al.}  Near-optimal single-photon sources in the solid
  state. \emph{Nature Photonics} \textbf{2016}, \emph{10}, 340--345\relax
\mciteBstWouldAddEndPuncttrue
\mciteSetBstMidEndSepPunct{\mcitedefaultmidpunct}
{\mcitedefaultendpunct}{\mcitedefaultseppunct}\relax
\EndOfBibitem
\bibitem[Schweickert \latin{et~al.}(2018)Schweickert, Jöns, Zeuner, Covre~da
  Silva, Huang, Lettner, Reindl, Zichi, Trotta, Rastelli, and
  Zwiller]{Schweickert_AppPhyLette2017}
Schweickert,~L.; Jöns,~K.~D.; Zeuner,~K.~D.; Covre~da Silva,~S.~F.; Huang,~H.;
  Lettner,~T.; Reindl,~M.; Zichi,~J.; Trotta,~R.; Rastelli,~A. \latin{et~al.}
  On-demand generation of background-free single photons from a solid-state
  source. \emph{Applied Physics Letters} \textbf{2018}, \emph{112},
  093106\relax
\mciteBstWouldAddEndPuncttrue
\mciteSetBstMidEndSepPunct{\mcitedefaultmidpunct}
{\mcitedefaultendpunct}{\mcitedefaultseppunct}\relax
\EndOfBibitem
\bibitem[Ding \latin{et~al.}(2016)Ding, He, Duan, Gregersen, Chen, Unsleber,
  Maier, Schneider, Kamp, H\"ofling, Lu, and Pan]{Ding_PRL2016}
Ding,~X.; He,~Y.; Duan,~Z.-C.; Gregersen,~N.; Chen,~M.-C.; Unsleber,~S.;
  Maier,~S.; Schneider,~C.; Kamp,~M.; H\"ofling,~S. \latin{et~al.}  On-Demand
  Single Photons with High Extraction Efficiency and Near-Unity
  Indistinguishability from a Resonantly Driven Quantum Dot in a Micropillar.
  \emph{Phys. Rev. Lett.} \textbf{2016}, \emph{116}, 020401\relax
\mciteBstWouldAddEndPuncttrue
\mciteSetBstMidEndSepPunct{\mcitedefaultmidpunct}
{\mcitedefaultendpunct}{\mcitedefaultseppunct}\relax
\EndOfBibitem
\bibitem[Schulte \latin{et~al.}(2015)Schulte, Hansom, Jones, Matthiesen,
  Le~Gall, and Atatüre]{Schulte_Science2015}
Schulte,~C.; Hansom,~J.; Jones,~A.; Matthiesen,~C.; Le~Gall,~C.; Atatüre,~M.
  Quadrature squeezed photons from a two-level system. \emph{Nature}
  \textbf{2015}, \emph{525}, 222--225\relax
\mciteBstWouldAddEndPuncttrue
\mciteSetBstMidEndSepPunct{\mcitedefaultmidpunct}
{\mcitedefaultendpunct}{\mcitedefaultseppunct}\relax
\EndOfBibitem
\bibitem[Schlehahn \latin{et~al.}(2016)Schlehahn, Thoma, Munnelly, Kamp,
  Höfling, Heindel, Schneider, and Reitzenstein]{Schlehahn_APL2016}
Schlehahn,~A.; Thoma,~A.; Munnelly,~P.; Kamp,~M.; Höfling,~S.; Heindel,~T.;
  Schneider,~C.; Reitzenstein,~S. An electrically driven cavity-enhanced source
  of indistinguishable photons with 61\% overall efficiency. \emph{APL
  Photonics} \textbf{2016}, \emph{1}, 011301\relax
\mciteBstWouldAddEndPuncttrue
\mciteSetBstMidEndSepPunct{\mcitedefaultmidpunct}
{\mcitedefaultendpunct}{\mcitedefaultseppunct}\relax
\EndOfBibitem
\bibitem[Aharonovich \latin{et~al.}(2016)Aharonovich, Englund, and
  Toth]{Aharonovich_NatPhoton2016}
Aharonovich,~I.; Englund,~D.; Toth,~M. Solid-state single-photon emitters.
  \emph{Nature Photon} \textbf{2016}, \emph{10}, 631–641\relax
\mciteBstWouldAddEndPuncttrue
\mciteSetBstMidEndSepPunct{\mcitedefaultmidpunct}
{\mcitedefaultendpunct}{\mcitedefaultseppunct}\relax
\EndOfBibitem
\bibitem[Sapienza \latin{et~al.}(2015)Sapienza, Davanço, Badolato, and
  Srinivasan]{Sapienza_Natcom2015}
Sapienza,~L.; Davanço,~M.; Badolato,~A.; Srinivasan,~K. Nanoscale optical
  positioning of single quantum dots for bright and pure single-photon
  emission. \emph{Nature Communications} \textbf{2015}, \emph{6}, 7833\relax
\mciteBstWouldAddEndPuncttrue
\mciteSetBstMidEndSepPunct{\mcitedefaultmidpunct}
{\mcitedefaultendpunct}{\mcitedefaultseppunct}\relax
\EndOfBibitem
\bibitem[Gschrey \latin{et~al.}(2013)Gschrey, Gericke, Schüßler, Schmidt,
  Schulze, Heindel, Rodt, Strittmatter, and Reitzenstein]{Gschrey_APL2013}
Gschrey,~M.; Gericke,~F.; Schüßler,~A.; Schmidt,~R.; Schulze,~J.-H.;
  Heindel,~T.; Rodt,~S.; Strittmatter,~A.; Reitzenstein,~S. In situ
  electron-beam lithography of deterministic single-quantum-dot mesa-structures
  using low-temperature cathodoluminescence spectroscopy. \emph{Applied Physics
  Letters} \textbf{2013}, \emph{102}, 251113\relax
\mciteBstWouldAddEndPuncttrue
\mciteSetBstMidEndSepPunct{\mcitedefaultmidpunct}
{\mcitedefaultendpunct}{\mcitedefaultseppunct}\relax
\EndOfBibitem
\bibitem[Senellart \latin{et~al.}(2017)Senellart, Solomon, and
  White]{Senellart_NatNano2017}
Senellart,~P.; Solomon,~G.; White,~A. High-performance semiconductor
  quantum-dot single-photon sources. \emph{Nature Nanotechnology}
  \textbf{2017}, \emph{12}, 1026--1039\relax
\mciteBstWouldAddEndPuncttrue
\mciteSetBstMidEndSepPunct{\mcitedefaultmidpunct}
{\mcitedefaultendpunct}{\mcitedefaultseppunct}\relax
\EndOfBibitem
\bibitem[BROWN and TWISS(1956)BROWN, and TWISS]{BROWN1956}
BROWN,~R.~H.; TWISS,~R.~Q. Correlation between Photons in two Coherent Beams of
  Light. \emph{Nature} \textbf{1956}, \emph{177}, 27--29\relax
\mciteBstWouldAddEndPuncttrue
\mciteSetBstMidEndSepPunct{\mcitedefaultmidpunct}
{\mcitedefaultendpunct}{\mcitedefaultseppunct}\relax
\EndOfBibitem
\bibitem[Migdall \latin{et~al.}(2013)Migdall, Polyakov, Fan, and
  Bienfang]{Migdallbook}
Migdall,~A.; Polyakov,~S.; Fan,~J.; Bienfang,~J. \emph{Single-Photon Generation
  and Detection, Volume 45}; Elsevier, 2013; pp 25--68\relax
\mciteBstWouldAddEndPuncttrue
\mciteSetBstMidEndSepPunct{\mcitedefaultmidpunct}
{\mcitedefaultendpunct}{\mcitedefaultseppunct}\relax
\EndOfBibitem
\bibitem[Hong \latin{et~al.}(1987)Hong, Ou, and Mandel]{HOM1987}
Hong,~C.~K.; Ou,~Z.~Y.; Mandel,~L. Measurement of subpicosecond time intervals
  between two photons by interference. \emph{Phys. Rev. Lett.} \textbf{1987},
  \emph{59}, 2044--2046\relax
\mciteBstWouldAddEndPuncttrue
\mciteSetBstMidEndSepPunct{\mcitedefaultmidpunct}
{\mcitedefaultendpunct}{\mcitedefaultseppunct}\relax
\EndOfBibitem
\bibitem[Di~Martino \latin{et~al.}(2014)Di~Martino, Sonnefraud, Tame,
  K\'ena-Cohen, Dieleman, \"Ozdemir, Kim, and Maier]{Martino_PRApp2014}
Di~Martino,~G.; Sonnefraud,~Y.; Tame,~M.~S.; K\'ena-Cohen,~S.; Dieleman,~F.;
  \"Ozdemir,~{\c{S}}.~K.; Kim,~M.~S.; Maier,~S.~A. Observation of Quantum
  Interference in the Plasmonic Hong-Ou-Mandel Effect. \emph{Phys. Rev.
  Applied} \textbf{2014}, \emph{1}, 034004\relax
\mciteBstWouldAddEndPuncttrue
\mciteSetBstMidEndSepPunct{\mcitedefaultmidpunct}
{\mcitedefaultendpunct}{\mcitedefaultseppunct}\relax
\EndOfBibitem
\bibitem[Broome \latin{et~al.}(2013)Broome, Fedrizzi, Rahimi-Keshari, Dove,
  Aaronson, Ralph, and White]{Broome_Science2013}
Broome,~M.~A.; Fedrizzi,~A.; Rahimi-Keshari,~S.; Dove,~J.; Aaronson,~S.;
  Ralph,~T.~C.; White,~A.~G. Photonic Boson Sampling in a Tunable Circuit.
  \emph{Science} \textbf{2013}, \emph{339}, 794--798\relax
\mciteBstWouldAddEndPuncttrue
\mciteSetBstMidEndSepPunct{\mcitedefaultmidpunct}
{\mcitedefaultendpunct}{\mcitedefaultseppunct}\relax
\EndOfBibitem
\bibitem[Spagnolo \latin{et~al.}(2014)Spagnolo, Vitelli, Bentivegna, Brod,
  Crespi, Flamini, Giacomini, Milani, Ramponi, Mataloni, Osellame, Galvão, and
  Sciarrino]{Spagnolo_Natph2014}
Spagnolo,~N.; Vitelli,~C.; Bentivegna,~M.; Brod,~D.~J.; Crespi,~A.;
  Flamini,~F.; Giacomini,~S.; Milani,~G.; Ramponi,~R.; Mataloni,~P.
  \latin{et~al.}  Experimental validation of photonic boson sampling.
  \emph{Nature Photonics} \textbf{2014}, \emph{8}, 615--620\relax
\mciteBstWouldAddEndPuncttrue
\mciteSetBstMidEndSepPunct{\mcitedefaultmidpunct}
{\mcitedefaultendpunct}{\mcitedefaultseppunct}\relax
\EndOfBibitem
\bibitem[Protesescu \latin{et~al.}(2015)Protesescu, Yakunin, Bodnarchuk, Krieg,
  Caputo, Hendon, Yang, Walsh, and Kovalenko]{Loredana_NanoLetters2015}
Protesescu,~L.; Yakunin,~S.; Bodnarchuk,~M.~I.; Krieg,~F.; Caputo,~R.;
  Hendon,~C.~H.; Yang,~R.~X.; Walsh,~A.; Kovalenko,~M.~V. Nanocrystals of
  Cesium Lead Halide Perovskites ($CsPbX_{3}$, X = Cl, Br, and I): Novel
  Optoelectronic Materials Showing Bright Emission with Wide Color Gamut.
  \emph{Nano Letters} \textbf{2015}, \emph{15}, 3692--3696\relax
\mciteBstWouldAddEndPuncttrue
\mciteSetBstMidEndSepPunct{\mcitedefaultmidpunct}
{\mcitedefaultendpunct}{\mcitedefaultseppunct}\relax
\EndOfBibitem
\bibitem[Winik \latin{et~al.}(2017)Winik, Cogan, Don, Schwartz, Gantz,
  Schmidgall, Livneh, Rapaport, Buks, and Gershoni]{Winik_PRB2017}
Winik,~R.; Cogan,~D.; Don,~Y.; Schwartz,~I.; Gantz,~L.; Schmidgall,~E.~R.;
  Livneh,~N.; Rapaport,~R.; Buks,~E.; Gershoni,~D. On-demand source of
  maximally entangled photon pairs using the biexciton-exciton radiative
  cascade. \emph{Phys. Rev. B} \textbf{2017}, \emph{95}, 235435\relax
\mciteBstWouldAddEndPuncttrue
\mciteSetBstMidEndSepPunct{\mcitedefaultmidpunct}
{\mcitedefaultendpunct}{\mcitedefaultseppunct}\relax
\EndOfBibitem
\bibitem[Trotta \latin{et~al.}(2014)Trotta, Wildmann, Zallo, Schmidt, and
  Rastelli]{Trotta_Nanolett2014}
Trotta,~R.; Wildmann,~J.~S.; Zallo,~E.; Schmidt,~O.~G.; Rastelli,~A. Highly
  Entangled Photons from Hybrid Piezoelectric-Semiconductor Quantum Dot
  Devices. \emph{Nano Letters} \textbf{2014}, \emph{14}, 3439--3444\relax
\mciteBstWouldAddEndPuncttrue
\mciteSetBstMidEndSepPunct{\mcitedefaultmidpunct}
{\mcitedefaultendpunct}{\mcitedefaultseppunct}\relax
\EndOfBibitem
\bibitem[Keil \latin{et~al.}(2017)Keil, Zopf, Chen, H\"{o}fer, Zhang, Schmidt,
  Ding, and Schmidt]{Solid-state_Nat2017}
Keil,~R.; Zopf,~M.; Chen,~Y.; H\"{o}fer,~B.; Zhang,~J.; Schmidt,~O.~G.;
  Ding,~F.; Schmidt,~O.~G. Solid-state ensemble of highly entangled photon
  sources at rubidium atomic transitions. \emph{Nature Communications}
  \textbf{2017}, \emph{8}, 15501\relax
\mciteBstWouldAddEndPuncttrue
\mciteSetBstMidEndSepPunct{\mcitedefaultmidpunct}
{\mcitedefaultendpunct}{\mcitedefaultseppunct}\relax
\EndOfBibitem
\bibitem[Huber \latin{et~al.}(2017)Huber, Reindl, Huo, Huang, Wildmann,
  Schmidt, Rastelli, and Trotta]{HighlyIndis_Nat2017}
Huber,~D.; Reindl,~M.; Huo,~Y.; Huang,~H.; Wildmann,~J.~S.; Schmidt,~O.~G.;
  Rastelli,~A.; Trotta,~R. Highly indistinguishable and strongly entangled
  photons from symmetric GaAs quantum dots. \emph{Nature Communications}
  \textbf{2017}, \emph{8}, 15506\relax
\mciteBstWouldAddEndPuncttrue
\mciteSetBstMidEndSepPunct{\mcitedefaultmidpunct}
{\mcitedefaultendpunct}{\mcitedefaultseppunct}\relax
\EndOfBibitem
\bibitem[Edamatsu(2007)]{EP_Generation_2007}
Edamatsu,~K. Entangled Photons: Generation, Observation, and Characterization.
  \emph{Japan Society of Applied Physics} \textbf{2007}, \emph{46},
  7175--7187\relax
\mciteBstWouldAddEndPuncttrue
\mciteSetBstMidEndSepPunct{\mcitedefaultmidpunct}
{\mcitedefaultendpunct}{\mcitedefaultseppunct}\relax
\EndOfBibitem
\bibitem[Moradi \latin{et~al.}(2017)Moradi, Harouni, and
  Naderi]{HighlyEPpairs_PRA2017}
Moradi,~T.; Harouni,~M.~B.; Naderi,~M.~H. Highly entangled photon pairs
  generated from the biexciton cascade transition in a
  quantum-dot--metal-nanoparticle hybrid system. \emph{Phys. Rev. A}
  \textbf{2017}, \emph{96}, 023836\relax
\mciteBstWouldAddEndPuncttrue
\mciteSetBstMidEndSepPunct{\mcitedefaultmidpunct}
{\mcitedefaultendpunct}{\mcitedefaultseppunct}\relax
\EndOfBibitem
\bibitem[Benson \latin{et~al.}(2000)Benson, Santori, Pelton, and
  Yamamoto]{Cascade_Semicon_PRL2000}
Benson,~O.; Santori,~C.; Pelton,~M.; Yamamoto,~Y. Regulated and Entangled
  Photons from a Single Quantum Dot. \emph{Phys. Rev. Lett.} \textbf{2000},
  \emph{84}, 2513--2516\relax
\mciteBstWouldAddEndPuncttrue
\mciteSetBstMidEndSepPunct{\mcitedefaultmidpunct}
{\mcitedefaultendpunct}{\mcitedefaultseppunct}\relax
\EndOfBibitem
\bibitem[Santori \latin{et~al.}(2002)Santori, Fattal, Pelton, Solomon, and
  Yamamoto]{Polarization-corre_PRB2002}
Santori,~C.; Fattal,~D.; Pelton,~M.; Solomon,~G.~S.; Yamamoto,~Y.
  Polarization-correlated photon pairs from a single quantum dot. \emph{Phys.
  Rev. B} \textbf{2002}, \emph{66}, 045308\relax
\mciteBstWouldAddEndPuncttrue
\mciteSetBstMidEndSepPunct{\mcitedefaultmidpunct}
{\mcitedefaultendpunct}{\mcitedefaultseppunct}\relax
\EndOfBibitem
\bibitem[Stace \latin{et~al.}(2003)Stace, Milburn, and
  Barnes]{EntangledBiexciton_PRB2003}
Stace,~T.~M.; Milburn,~G.~J.; Barnes,~C. H.~W. Entangled two-photon source
  using biexciton emission of an asymmetric quantum dot in a cavity.
  \emph{Phys. Rev. B} \textbf{2003}, \emph{67}, 085317\relax
\mciteBstWouldAddEndPuncttrue
\mciteSetBstMidEndSepPunct{\mcitedefaultmidpunct}
{\mcitedefaultendpunct}{\mcitedefaultseppunct}\relax
\EndOfBibitem
\bibitem[Stevenson \latin{et~al.}(2006)Stevenson, Young, Atkinson, Cooper, A.,
  and Shields]{CascadeSemiconNat2006}
Stevenson,~R.~M.; Young,~R.~J.; Atkinson,~P.; Cooper,~K.; A.,~R.~D.;
  Shields,~A.~J. A semiconductor source of triggered entangled photon pairs.
  \emph{Nature} \textbf{2006}, \emph{439}, 179–182\relax
\mciteBstWouldAddEndPuncttrue
\mciteSetBstMidEndSepPunct{\mcitedefaultmidpunct}
{\mcitedefaultendpunct}{\mcitedefaultseppunct}\relax
\EndOfBibitem
\bibitem[Gerardot \latin{et~al.}(2007)Gerardot, Seidl, Dalgarno, Warburton,
  Granados, Garcia, Kowalik, Krebs, Karrai, Badolato, and
  Petroff]{ManipulatingFSS_APL2007}
Gerardot,~B.~D.; Seidl,~S.; Dalgarno,~P.~A.; Warburton,~R.~J.; Granados,~D.;
  Garcia,~J.~M.; Kowalik,~K.; Krebs,~O.; Karrai,~K.; Badolato,~A.
  \latin{et~al.}  Manipulating exciton fine structure in quantum dots with a
  lateral electric field. \emph{Applied Physics Letters} \textbf{2007},
  \emph{90}, 041101\relax
\mciteBstWouldAddEndPuncttrue
\mciteSetBstMidEndSepPunct{\mcitedefaultmidpunct}
{\mcitedefaultendpunct}{\mcitedefaultseppunct}\relax
\EndOfBibitem
\bibitem[Singh and Bester(2010)Singh, and Bester]{LowerBoundPRL2010}
Singh,~R.; Bester,~G. Lower Bound for the Excitonic Fine Structure Splitting in
  Self-Assembled Quantum Dots. \emph{Phys. Rev. Lett.} \textbf{2010},
  \emph{104}, 196803\relax
\mciteBstWouldAddEndPuncttrue
\mciteSetBstMidEndSepPunct{\mcitedefaultmidpunct}
{\mcitedefaultendpunct}{\mcitedefaultseppunct}\relax
\EndOfBibitem
\bibitem[Akopian \latin{et~al.}(2006)Akopian, Lindner, Poem, Berlatzky, Avron,
  Gershoni, Gerardot, and Petroff]{Entangled_AkopianPRL2006}
Akopian,~N.; Lindner,~N.~H.; Poem,~E.; Berlatzky,~Y.; Avron,~J.; Gershoni,~D.;
  Gerardot,~B.~D.; Petroff,~P.~M. Entangled Photon Pairs from Semiconductor
  Quantum Dots. \emph{Phys. Rev. Lett.} \textbf{2006}, \emph{96}, 130501\relax
\mciteBstWouldAddEndPuncttrue
\mciteSetBstMidEndSepPunct{\mcitedefaultmidpunct}
{\mcitedefaultendpunct}{\mcitedefaultseppunct}\relax
\EndOfBibitem
\bibitem[Pathak and Hughes(2009)Pathak, and Hughes]{Cavity-assisted_PRB2009}
Pathak,~P.~K.; Hughes,~S. Cavity-assisted fast generation of entangled photon
  pairs through the biexciton-exciton cascade. \emph{Phys. Rev. B}
  \textbf{2009}, \emph{80}, 155325\relax
\mciteBstWouldAddEndPuncttrue
\mciteSetBstMidEndSepPunct{\mcitedefaultmidpunct}
{\mcitedefaultendpunct}{\mcitedefaultseppunct}\relax
\EndOfBibitem
\bibitem[Pathak and Hughes(2009)Pathak, and Hughes]{Pathak_PRB2009}
Pathak,~P.~K.; Hughes,~S. Generation of entangled photon pairs from a single
  quantum dot embedded in a planar photonic-crystal cavity. \emph{Phys. Rev. B}
  \textbf{2009}, \emph{79}, 205416\relax
\mciteBstWouldAddEndPuncttrue
\mciteSetBstMidEndSepPunct{\mcitedefaultmidpunct}
{\mcitedefaultendpunct}{\mcitedefaultseppunct}\relax
\EndOfBibitem
\bibitem[Scully and Zubairy(1997)Scully, and Zubairy]{Scullybook}
Scully,~M.; Zubairy,~M.~S. \emph{Quantum Optics}; Cambridge University Press,
  Cambridge, UK, 1997\relax
\mciteBstWouldAddEndPuncttrue
\mciteSetBstMidEndSepPunct{\mcitedefaultmidpunct}
{\mcitedefaultendpunct}{\mcitedefaultseppunct}\relax
\EndOfBibitem
\bibitem[Avanaki and Schatz(2019)Avanaki, and Schatz]{AvanakiJPCL2019}
Avanaki,~K.~N.; Schatz,~G.~C. Entangled Photon Resonance Energy Transfer in
  Arbitrary Media. \emph{The Journal of Physical Chemistry Letters}
  \textbf{2019}, \emph{10}, 3181--3188\relax
\mciteBstWouldAddEndPuncttrue
\mciteSetBstMidEndSepPunct{\mcitedefaultmidpunct}
{\mcitedefaultendpunct}{\mcitedefaultseppunct}\relax
\EndOfBibitem
\bibitem[Loudon(1963)]{Loudonbook}
Loudon,~R. \emph{The Quantum Theory of Light}; Clarendon Press, Oxford,
  1963\relax
\mciteBstWouldAddEndPuncttrue
\mciteSetBstMidEndSepPunct{\mcitedefaultmidpunct}
{\mcitedefaultendpunct}{\mcitedefaultseppunct}\relax
\EndOfBibitem
\bibitem[Morisue \latin{et~al.}(2018)Morisue, Omagari, Ueno, Nakanishi,
  Hasegawa, Yamamoto, Matsui, Sasaki, Hikima, and Sakurai]{2018acsomega}
Morisue,~M.; Omagari,~S.; Ueno,~I.; Nakanishi,~T.; Hasegawa,~Y.; Yamamoto,~S.;
  Matsui,~J.; Sasaki,~S.; Hikima,~T.; Sakurai,~S. Fully Conjugated Porphyrin
  Glass: Collective Light-Harvesting Antenna for Near-Infrared Fluorescence
  beyond 1 $\mu m$. \emph{ACS Omega} \textbf{2018}, \emph{3}, 4466--4474\relax
\mciteBstWouldAddEndPuncttrue
\mciteSetBstMidEndSepPunct{\mcitedefaultmidpunct}
{\mcitedefaultendpunct}{\mcitedefaultseppunct}\relax
\EndOfBibitem
\bibitem[Young \latin{et~al.}(2006)Young, Stevenson, Atkinson, Cooper, Ritchie,
  and Shields]{CascadeSemiconNJPhys2006}
Young,~R.~J.; Stevenson,~R.~M.; Atkinson,~P.; Cooper,~K.; Ritchie,~D.~A.;
  Shields,~A.~J. Improved fidelity of triggered entangled photons from single
  quantum dots. \emph{New Journal of Physics} \textbf{2006}, \emph{8}, 29\relax
\mciteBstWouldAddEndPuncttrue
\mciteSetBstMidEndSepPunct{\mcitedefaultmidpunct}
{\mcitedefaultendpunct}{\mcitedefaultseppunct}\relax
\EndOfBibitem
\bibitem[Coles \latin{et~al.}(2014)Coles, Somaschi, Michetti, Clark,
  Lagoudakis, Savvidis, Lidzey, and and]{Polariton-mediatedNat2014}
Coles,~D.~M.; Somaschi,~N.; Michetti,~P.; Clark,~C.; Lagoudakis,~P.~G.;
  Savvidis,~P.~G.; Lidzey,~D.~G.; and, Polariton-mediated energy transfer
  between organic dyes in a strongly coupled optical microcavity. \emph{Nature
  Materials} \textbf{2014}, \emph{13}, 712\relax
\mciteBstWouldAddEndPuncttrue
\mciteSetBstMidEndSepPunct{\mcitedefaultmidpunct}
{\mcitedefaultendpunct}{\mcitedefaultseppunct}\relax
\EndOfBibitem
\bibitem[Zhong \latin{et~al.}(2016)Zhong, Chervy, Wang, George, Thomas,
  Hutchison, Devaux, Genet, and Ebbesen]{Ebbesen_Angewandte2016}
Zhong,~X.; Chervy,~T.; Wang,~S.; George,~J.; Thomas,~A.; Hutchison,~J.~A.;
  Devaux,~E.; Genet,~C.; Ebbesen,~T.~W. Non-Radiative Energy Transfer Mediated
  by Hybrid Light-Matter States. \emph{Angewandte Chemie International Edition}
  \textbf{2016}, \emph{55}, 6202--6206\relax
\mciteBstWouldAddEndPuncttrue
\mciteSetBstMidEndSepPunct{\mcitedefaultmidpunct}
{\mcitedefaultendpunct}{\mcitedefaultseppunct}\relax
\EndOfBibitem
\bibitem[Zhong \latin{et~al.}(2017)Zhong, Chervy, Zhang, Thomas, George, Genet,
  Hutchison, and Ebbesen]{Ebbesen_Angewandte2017}
Zhong,~X.; Chervy,~T.; Zhang,~L.; Thomas,~A.; George,~J.; Genet,~C.;
  Hutchison,~J.~A.; Ebbesen,~T.~W. Energy Transfer between Spatially Separated
  Entangled Molecules. \emph{Angewandte Chemie International Edition}
  \textbf{2017}, \emph{56}, 9034--9038\relax
\mciteBstWouldAddEndPuncttrue
\mciteSetBstMidEndSepPunct{\mcitedefaultmidpunct}
{\mcitedefaultendpunct}{\mcitedefaultseppunct}\relax
\EndOfBibitem
\bibitem[Ripka \latin{et~al.}(2018)Ripka, K{\"u}bler, L{\"o}w, and
  Pfau]{Ripka_science2018}
Ripka,~F.; K{\"u}bler,~H.; L{\"o}w,~R.; Pfau,~T. A room-temperature
  single-photon source based on strongly interacting Rydberg atoms.
  \emph{Science} \textbf{2018}, \emph{362}, 446--449\relax
\mciteBstWouldAddEndPuncttrue
\mciteSetBstMidEndSepPunct{\mcitedefaultmidpunct}
{\mcitedefaultendpunct}{\mcitedefaultseppunct}\relax
\EndOfBibitem
\bibitem[Crocker \latin{et~al.}(2019)Crocker, Lichtman, Sosnova, Carter,
  Scarano, and Monroe]{Crocker_Express2019}
Crocker,~C.; Lichtman,~M.; Sosnova,~K.; Carter,~A.; Scarano,~S.; Monroe,~C.
  High purity single photons entangled with an atomic qubit. \emph{Opt.
  Express} \textbf{2019}, \emph{27}, 28143--28149\relax
\mciteBstWouldAddEndPuncttrue
\mciteSetBstMidEndSepPunct{\mcitedefaultmidpunct}
{\mcitedefaultendpunct}{\mcitedefaultseppunct}\relax
\EndOfBibitem
\bibitem[Silva \latin{et~al.}(2016)Silva, Sánchez~Muñoz, Ballarini,
  González-Tudela, de~Giorgi, Gigli, West, Pfeiffer, del Valle, Sanvitto, and
  Laussy]{Silva_SciRep2016}
Silva,~B.; Sánchez~Muñoz,~C.; Ballarini,~D.; González-Tudela,~A.;
  de~Giorgi,~M.; Gigli,~G.; West,~K.; Pfeiffer,~L.; del Valle,~E.; Sanvitto,~D.
  \latin{et~al.}  The colored Hanbury Brown–Twiss effect. \emph{Scientific
  Reports} \textbf{2016}, \emph{6}, 37980\relax
\mciteBstWouldAddEndPuncttrue
\mciteSetBstMidEndSepPunct{\mcitedefaultmidpunct}
{\mcitedefaultendpunct}{\mcitedefaultseppunct}\relax
\EndOfBibitem
\bibitem[Kiraz \latin{et~al.}(2004)Kiraz, Atat\"ure, and
  Imamo\ifmmode~\breve{g}\else \u{g}\fi{}lu]{Kiraz_PRA2004}
Kiraz,~A.; Atat\"ure,~M.; Imamo\ifmmode~\breve{g}\else \u{g}\fi{}lu,~A.
  Quantum-dot single-photon sources: Prospects for applications in linear
  optics quantum-information processing. \emph{Phys. Rev. A} \textbf{2004},
  \emph{69}, 032305\relax
\mciteBstWouldAddEndPuncttrue
\mciteSetBstMidEndSepPunct{\mcitedefaultmidpunct}
{\mcitedefaultendpunct}{\mcitedefaultseppunct}\relax
\EndOfBibitem
\bibitem[Ahn(2020)]{Ahn_OptExpress2020}
Ahn,~K.~J. Temporal dynamics of zero-delay second order correlation function
  and spectral entanglement of two photons emitted from ladder-type atomic
  three-level systems. \emph{Opt. Express} \textbf{2020}, \emph{28},
  1790--1804\relax
\mciteBstWouldAddEndPuncttrue
\mciteSetBstMidEndSepPunct{\mcitedefaultmidpunct}
{\mcitedefaultendpunct}{\mcitedefaultseppunct}\relax
\EndOfBibitem
\bibitem[Hamsen \latin{et~al.}(2018)Hamsen, Tolazzi, Wilk, and
  Rempe]{Hamsen_Natphys2018}
Hamsen,~C.; Tolazzi,~K.~N.; Wilk,~T.; Rempe,~G. Strong coupling between photons
  of two light fields mediated by one atom. \emph{Nature Physics}
  \textbf{2018}, \emph{14}, 885--889\relax
\mciteBstWouldAddEndPuncttrue
\mciteSetBstMidEndSepPunct{\mcitedefaultmidpunct}
{\mcitedefaultendpunct}{\mcitedefaultseppunct}\relax
\EndOfBibitem
\bibitem[He \latin{et~al.}(2016)He, Iff, Lundt, Baumann, Davanco, Srinivasan,
  Höfling, and Schneider]{Yu-Ming_Nat2016}
He,~Y.-M.; Iff,~O.; Lundt,~N.; Baumann,~V.; Davanco,~M.; Srinivasan,~K.;
  Höfling,~S.; Schneider,~C. Cascaded emission of single photons from the
  biexciton in monolayered $WSe_{2}$. \emph{Nature Communications}
  \textbf{2016}, \emph{7}, 13409\relax
\mciteBstWouldAddEndPuncttrue
\mciteSetBstMidEndSepPunct{\mcitedefaultmidpunct}
{\mcitedefaultendpunct}{\mcitedefaultseppunct}\relax
\EndOfBibitem
\bibitem[Law \latin{et~al.}(2000)Law, Walmsley, and Eberly]{Eberly_PRL2000}
Law,~C.~K.; Walmsley,~I.~A.; Eberly,~J.~H. Continuous Frequency Entanglement:
  Effective Finite Hilbert Space and Entropy Control. \emph{Phys. Rev. Lett.}
  \textbf{2000}, \emph{84}, 5304--5307\relax
\mciteBstWouldAddEndPuncttrue
\mciteSetBstMidEndSepPunct{\mcitedefaultmidpunct}
{\mcitedefaultendpunct}{\mcitedefaultseppunct}\relax
\EndOfBibitem
\bibitem[Eberly(2006)]{Eberly2006}
Eberly,~J.~H. Schmidt analysis of pure-state entanglement. \emph{Laser Physics}
  \textbf{2006}, \emph{16}, 921--926\relax
\mciteBstWouldAddEndPuncttrue
\mciteSetBstMidEndSepPunct{\mcitedefaultmidpunct}
{\mcitedefaultendpunct}{\mcitedefaultseppunct}\relax
\EndOfBibitem
\bibitem[Chen \latin{et~al.}(2017)Chen, Bo, Niu, Xu, Zhang, Shapiro, and
  Wong]{Chen2017}
Chen,~C.; Bo,~C.; Niu,~M.~Y.; Xu,~F.; Zhang,~Z.; Shapiro,~J.~H.; Wong,~F. N.~C.
  Efficient generation and characterization of spectrally factorable biphotons.
  \emph{Opt. Express} \textbf{2017}, \emph{25}, 7300--7312\relax
\mciteBstWouldAddEndPuncttrue
\mciteSetBstMidEndSepPunct{\mcitedefaultmidpunct}
{\mcitedefaultendpunct}{\mcitedefaultseppunct}\relax
\EndOfBibitem
\bibitem[Lamata and Le{\'{o}}n(2005)Lamata, and Le{\'{o}}n]{Lamata_2005}
Lamata,~L.; Le{\'{o}}n,~J. Dealing with entanglement of continuous variables:
  Schmidt decomposition with discrete sets of orthogonal functions.
  \emph{Journal of Optics B: Quantum and Semiclassical Optics} \textbf{2005},
  \emph{7}, 224--229\relax
\mciteBstWouldAddEndPuncttrue
\mciteSetBstMidEndSepPunct{\mcitedefaultmidpunct}
{\mcitedefaultendpunct}{\mcitedefaultseppunct}\relax
\EndOfBibitem
\bibitem[Valencia \latin{et~al.}(2007)Valencia, Cer\'e, Shi, Molina-Terriza,
  and Torres]{Valencia_PRL2007}
Valencia,~A.; Cer\'e,~A.; Shi,~X.; Molina-Terriza,~G.; Torres,~J.~P. Shaping
  the Waveform of Entangled Photons. \emph{Phys. Rev. Lett.} \textbf{2007},
  \emph{99}, 243601\relax
\mciteBstWouldAddEndPuncttrue
\mciteSetBstMidEndSepPunct{\mcitedefaultmidpunct}
{\mcitedefaultendpunct}{\mcitedefaultseppunct}\relax
\EndOfBibitem
\bibitem[Hudson \latin{et~al.}(2007)Hudson, Stevenson, Bennett, Young, Nicoll,
  Atkinson, Cooper, Ritchie, and Shields]{Hudson_PRL2007}
Hudson,~A.~J.; Stevenson,~R.~M.; Bennett,~A.~J.; Young,~R.~J.; Nicoll,~C.~A.;
  Atkinson,~P.; Cooper,~K.; Ritchie,~D.~A.; Shields,~A.~J. Coherence of an
  Entangled Exciton-Photon State. \emph{Phys. Rev. Lett.} \textbf{2007},
  \emph{99}, 266802\relax
\mciteBstWouldAddEndPuncttrue
\mciteSetBstMidEndSepPunct{\mcitedefaultmidpunct}
{\mcitedefaultendpunct}{\mcitedefaultseppunct}\relax
\EndOfBibitem
\bibitem[Stevenson \latin{et~al.}(2008)Stevenson, Hudson, Bennett, Young,
  Nicoll, Ritchie, and Shields]{Stevenson_PRL2008}
Stevenson,~R.~M.; Hudson,~A.~J.; Bennett,~A.~J.; Young,~R.~J.; Nicoll,~C.~A.;
  Ritchie,~D.~A.; Shields,~A.~J. Evolution of Entanglement Between
  Distinguishable Light States. \emph{Phys. Rev. Lett.} \textbf{2008},
  \emph{101}, 170501\relax
\mciteBstWouldAddEndPuncttrue
\mciteSetBstMidEndSepPunct{\mcitedefaultmidpunct}
{\mcitedefaultendpunct}{\mcitedefaultseppunct}\relax
\EndOfBibitem
\end{mcitethebibliography}

\end{document}